\documentstyle[11pt]{article}
\title{Momentum and Coordinate Space Three-nucleon Potentials}
\author{S.\ A.\ Coon
 \\Physics Department \\New Mexico State University\\
  Las Cruces, NM  88003, USA \and M.\ T.\ Pe\~{n}a \\
CEBAF, Theory Group \\12000 Jefferson Ave,\\Newport News, VA 23606,
USA\thanks{Permanent address: Centro de
F\'\i sica Nuclear,  1699 Lisboa Codex, Portugal.} .}

\newcommand{\fone} {\mbox{ $F_{1 \rho}(k^{2})\,$}}
\newcommand{\fonep} {\mbox{ $F_{1 \rho}(k'^{2})\,$}}
\newcommand{\ftwo} {\mbox{ $F_{2 \rho}(k^{2})\,$}}
\newcommand{\ftwop} {\mbox{ $F_{2 \rho}(k'^{2})\,$}}
\newcommand{\bigGstar}  {\mbox{  $G^{*}_{M\rho}$}}
\newcommand{\bigGstark}  {\mbox{ $
               G^{*}_{M\rho} F_{\rho N\Delta}(k^{2})\, $}}
\newcommand{\gstar}  {\mbox{ $ g^{*}F_{\pi N\Delta}(q^{2})\, $}}
\newcommand{\Fpinn}  {\mbox{ $F_{\pi NN}(q^{2})\, $}}
\newcommand{\Frhondel}  {\mbox{ $F_{\rho N\Delta}(k^{2})\, $}}
\newcommand{\Fpindel}  {\mbox{ $F_{\pi N\Delta}(q^{2})\, $}}
\newcommand{\kapparho} {\mbox {$ \kappa_{\rho}  $}}
\newcommand{\rhoNN} {\mbox {$ \vec{\sigma_{1}}\times\vec{k} $}}
\newcommand{\piNN}  {\mbox {$ \vec{\sigma_{2}}\cdot\vec{q}  $}}

\newcommand{\mompotrp} {\mbox { $
<\vec{p'_{1}},\vec{p'_{2}},\vec{p'_{3}}|{\cal W}^{-}_{\rho\pi}(3A)
|\vec{p_{1}},\vec{p_{2}},\vec{p_{3}}> $}}
\newcommand{\mompotrpdelplus} {\mbox { $
<\vec{p'_{1}},\vec{p'_{2}},\vec{p'_{3}}|{\cal W}^{+}_{\rho\pi\Delta}(3A)
|\vec{p_{1}},\vec{p_{2}},\vec{p_{3}}> $}}
\newcommand{\mompotrpdelminus} {\mbox { $
<\vec{p'_{1}},\vec{p'_{2}},\vec{p'_{3}}|{\cal W}^{-}_{\rho\pi\Delta}(3A)
|\vec{p_{1}},\vec{p_{2}},\vec{p_{3}}> $}}

\newcommand{\mompotrr} {\mbox { $
<\vec{p'_{1}},\vec{p'_{2}},\vec{p'_{3}}|{ W}^{-}_{TT}
|\vec{p_{1}},\vec{p_{2}},\vec{p_{3}}> $}}
\newcommand{\mompotrrplusdel} {\mbox { $
<\vec{p'_{1}},\vec{p'_{2}},\vec{p'_{3}}|{ W}^{+}_{\rho\rho\Delta}
|\vec{p_{1}},\vec{p_{2}},\vec{p_{3}}> $}}
\newcommand{\mompotrrminusdel} {\mbox { $
<\vec{p'_{1}},\vec{p'_{2}},\vec{p'_{3}}|{ W}^{-}_{\rho\rho\Delta}
|\vec{p_{1}},\vec{p_{2}},\vec{p_{3}}> $}}

\newcommand{\corpotrp} {\mbox { $
<\vec{r'_{1}},\vec{r'_{2}},\vec{r'_{3}}|{\cal W}^{-}_{\rho\pi}(3A)
|\vec{r_{1}},\vec{r_{2}},\vec{r_{3}}> $}}
\newcommand{\corpotrpdelplus} {\mbox { $
<\vec{r'_{1}},\vec{r'_{2}},\vec{r'_{3}}|{\cal W}^{+}_{\rho\pi\Delta}(3A)
|\vec{r_{1}},\vec{r_{2}},\vec{r_{3}}> $}}
\newcommand{\corpotrpdelminus} {\mbox { $
<\vec{r'_{1}},\vec{r'_{2}},\vec{r'_{3}}|{\cal W}^{-}_{\rho\pi\Delta}(3A)
|\vec{r_{1}},\vec{r_{2}},\vec{r_{3}}> $}}

\newcommand{\oddiso} {\mbox
       { $\vec{\tau_{1}}\cdot\vec{\tau_{2}}\times\vec{\tau_{3}} $}}
\newcommand{\eveniso} {\mbox
       { $\vec{\tau_{1}}\cdot\vec{\tau_{2}} $}}
\newcommand{\kfour} {\mbox {$ \vec{\sigma_{1}}\cdot\vec{q}
           \vec{\sigma_{2}}\cdot\vec{q} $}}
\newcommand{\kfive} {\mbox {$ \vec{\sigma_{1}}\cdot\vec{k}
           \vec{\sigma_{2}}\cdot\vec{q} $}}

\newcommand{\ksix} {\mbox {$ i\rhoNN\cdot
       \vec{\sigma_{3}}\piNN $}}
\newcommand{\kseven} {\mbox {$ i \rhoNN\cdot\vec{q}
       \vec{\sigma_{3}}\cdot\vec{k}\piNN $}}

\newcommand{\corpotrr} {\mbox { $
<\vec{r'_{1}},\vec{r'_{2}},\vec{r'_{3}}|{\cal W}^{(-)}_{3TT}
|\vec{r_{1}},\vec{r_{2}},\vec{r_{3}}> $}}
\newcommand{\corpotrrplusdel} {\mbox { $
<\vec{r'_{1}},\vec{r'_{2}},\vec{r'_{3}}|{\cal W}^{(+)}_{\rho\rho\Delta}
|\vec{r_{1}},\vec{r_{2}},\vec{r_{3}}> $}}
\newcommand{\corpotrrminusdel} {\mbox { $
<\vec{r'_{1}},\vec{r'_{2}},\vec{r'_{3}}|{\cal W}^{(-)}_{\rho\rho\Delta}
|\vec{r_{1}},\vec{r_{2}},\vec{r_{3}}> $}}

\newcommand{\rhocrosscross} {\mbox { $
((\vec{\sigma_{1}}\times\vec{k})\times\vec{k})\cdot
((\vec{\sigma_{2}}\times\vec{k'})\times\vec{k'})  $}}
\newcommand{\rhotriplecross} {\mbox { $
((\vec{\sigma_{1}}\times\vec{k})\times\vec{k})\times
((\vec{\sigma_{2}}\times\vec{k'})\times\vec{k'})  $}}

\begin{document}

\maketitle
\begin{abstract}
     In this paper we give explicit formulae in momentum and coordinate
space for the three-nucleon potentials due to $\rho$ and $\pi$ meson
exchange, derived from off-mass-shell meson-nucleon scattering
amplitudes which are constrained by the symmetries of QCD and by the
experimental data. Those potentials have already been applied to nuclear
matter calculations. Here we display additional terms which appear to be
the most important for nuclear structure. The potentials are decomposed
in a way that separates the contributions of different physical
mechanisms involved in the meson-nucleon amplitudes. The same type of
decomposition is presented for the $\pi - \pi$ TM force: the $\Delta$,
the chiral symmetry breaking and the nucleon pair terms are isolated.
\end{abstract}
\addtocounter{footnote}{-1}
\newpage

\baselineskip 0.6cm

\section{Introduction}

Three-nucleon potentials based on $\pi$ and $\rho$ meson exchange have
been derived from different underlying approaches.  The recent papers
which include $\rho$ exchange \cite{Mart,Robilotta,Ellis,Gari} tend to
extend to the rho meson whichever approach had been already used for the
two pion exchange potential considered earlier (because of its
longer range).  The early history of the $\pi - \pi$ three-nucleon
potential was categorized in this way and summarized in \cite{Coon1979}.
Those potentials built upon the excitation of the nucleon into a
$\Delta$ isobar were the first and contain the least physical input.
These $\Delta$ mediated potentials have recently been extended to
include $\rho$ exchange in Refs. \cite{Mart} and \cite{Gari}.  Soon
after the first $\Delta$ mediated $\pi - \pi$ potential was constructed,
it was realized that chiral symmetry (breaking) must be included in the
potential. That is, the theoretical $\pi$N amplitudes must satisfy
well-defined chiral (so called ``soft pion") limits \cite{BGG}.  One
way to accomplish this begins with the use of (effective)
Lagrangians with pions,
nucleons, deltas etc. which satisfy approximate chiral symmetry
\cite{Rob1,CF,Weinberg}. This approach has
recently been extended to include $\rho$ mesons, so that the assumed
Lagrangians must also obey gauge invariance \cite{Robilotta}. This
approach has built in the correct symmetries but the couplings must be
estimated and, furthermore, the results have not really been
tested against the pion-nucleon data \cite{Murphy}.

Alternatively, the approach used in the Tucson-Melbourne (TM) family of
forces is based upon applying the Ward identities of current algebra to
axial-vector nucleon scattering. The Ward identities are saturated with
nucleon and $\Delta(1230)$ poles. Then employing PCAC (partial
conservation of the axial-vector current), one can derive expressions
for the on-mass-shell pion-nucleon scattering amplitudes which map out
satisfactorily the empirical coefficients of the H\"{o}hler subthreshold
crossing symmetric expansion based on dispersion relations
\cite{hohlerbook}. The off-mass-shell extrapolation (needed for the
exchange of virtual, spacelike pions in a nuclear force diagram) of the most
important amplitude $\bar{F}^+$ can be written in a form which depends
on measured on-shell amplitudes only.  This rewriting of the amplitude
to bury all reference to models of the $\Delta$ exploits a convenient
correspondence between the structure of the terms corresponding to
spontaneously broken chiral symmetry and the structure of the model
$\Delta$ term. However, both the field theoretic \cite{OO} and dispersion
theoretic  \cite{ST} $\Delta$ contributions to $\bar{F}^+$ result in an
equally excellent description of the data \cite{sigff}, so the early
emphasis on maximum model independence can perhaps be relaxed. This will
be done in the present paper.

Extending this current algebra-PCAC program to the Compton-like
processes  \cite{MacMullen} $\gamma_{\nu} + N \rightarrow A_{\mu} + N$, where
$\gamma_{\nu}$ is a vector photon and $A_{\mu}$ is an axial-vector
current,  provides, through vector dominance,  a solid
basis for modeling a $\rho-\pi$ exchange three-nucleon force.  Again
the amplitude is obtained via Ward identities but this time exploiting
in addition the gauge condition on the off-shell electroproduction amplitude.
Unfortunately, there is no empirical subthreshold expansion of the
invariant amplitudes of pion photo- and electroproduction.  Instead, the
amplitudes are tested against the soft pion theorems and (on-shell)
against multipoles at threshold or slightly above threshold.  The
amplitude which forms the basis of the TM $\rho-\pi$ force meets these
tests \cite{MacMullen}, including the challenge of the recently measured
neutral pion photoproduction \cite{BPS}.

The so-called Tucson-Melbourne (TM) family of three-nucleon forces, in
the particular case of $\rho-\pi$ and $\rho-\rho$ exchange, has never
been displayed in a form suitable for applications in general nuclear
systems (in particular, we have in mind few-nucleon systems).
In the work of Ref. \cite{Ellis} those potentials were applied to
nuclear matter where certain approximations could be
carried out. As a result, Ref.  \cite{Ellis} failed to provide a
simultaneously complete and convenient explicit form of the TM $\rho-\pi$ and
$\rho-\rho$ forces, ready for immediate use in any calculation, as we
propose to present in this paper.

Nevertheless, the nuclear matter calculation of ref  \cite{Ellis} was
indicative of the role of the $\rho$ meson in screening the $\pi$
exchange effect. The same effect had already been seen in the nuclear
matter calculation of Ref. \cite{Mart} where a three-nucleon force
constructed only from $NN-N\Delta$ transition potentials was used.

Furthermore, calculations of the triton bound state with inclusion of
the TM 2$\pi$ exchange potential yield over-binding and strong
dependence on the $\pi NN$ form factor regulator mass \cite{Chen,Sas1}.
This reflects the presence in the $\pi-\pi$ TM force of a term (usually
denoted in the published literature as the ``c" term \cite{Coon1979}) in
part due to chiral symmetry breaking and responsible for the rapid
variation of the lower-order terms in the $\pi-N$ amplitude, as demanded
by the soft pion theorems. It appears then natural to extend the
calculations on the three-nucleon system to the inclusion of the TM
$\rho-\pi$  and $\rho-\rho$ forces hoping to cancel part of the
$\pi-\pi$ force effect, in a similar fashion to what happens in the
two-nucleon interactions. Moreover, since the leading terms of the
expansions of the the $\rho N \rightarrow \pi N$ and $\rho N \rightarrow
\rho N$ amplitudes do not have to build in a drastic variation in the
low energy region, they do not have such short range singularities. In
this spirit, the Sendai group included a $\rho-\pi$ force in a triton
calculation \cite{Sas2} which, however, due to lack of a consistent body
of information in the literature, mixes several terms and parameter
prescriptions of more than one origin.

Another motivation to have  $\rho-\pi$  and $\rho-\rho$ forces derived
from the same principles used in the construction of the TM $\pi-\pi $
force is to have available a more complete two-meson-exchange
three-nucleon interaction, which is needed
to draw conclusions about the role played by the delta-isobar in the
force. Only after introducing the consistent full family of forces in
the hamiltonian, does it become legitimate to make comparisons with the
Hannover (two-body $NN \rightarrow N\Delta$ transition potential)
approach and judge how much effect comes
from true $\Delta$-isobar propagation. A first study of this question,
restricted however to the use of the TM $\pi-\pi$  force, appeared most
recently in the literature \cite{Stadler1}. Another one is in
preparation where the $\rho-\pi$ and $\rho-\rho$ TM force input is
needed \cite{Stadler2,Stadler3}. In the context of these recent developments
in few-body calculations, the present paper supplies the necessary
information to address the one-decade unsolved problem of comparing and
constrasting the Hannover with the Tucson-Melbourne approaches.

Finally, and aiming beyond the triton bound state problem, the stage of
nowadays powerful calculations allows the information of this paper to
be applied to 1) the present concentrated effort on experimental
searches for direct evidence of three-nucleon forces in the
three-nucleon continuum guided by  the three nucleon-scattering
calculations of the Bochum group \cite{Witala}; and 2) the $\alpha$
particle wavefunction calculation \cite{Kamada}, under the strong indications
from existing calculations \cite{Sapporo,CoonZab,Carlson,Wiringa} that the
three-nucleon force effects, when fully considered, will be relevant in
bringing the binding energy closer to its experimental value.

\section{The $\pi - \pi$ Potentials}

In this section we carry out the program of splitting the $\pi-\pi$
force into the contributions relative to the different physical
processes underlying the $\pi$N amplitude. Historically the current
algebra amplitudes were given in a form which emphasized their model
independent character. That was achieved by not disentangling explicitely
the $\Delta$ contribution from the chiral symmetry breaking terms.
Such an attitude was
justified by allowing a more direct relation to empirical
quantities, where that separation could not be seen.
The process of undoing this way of displaying the force is implicit in
the set of initial papers \cite{Coon1979,CG}.
Here we merely summarize the procedure by showing the main steps.

The $\pi-\pi$ force was derived from a non-relativistic reduction
of the Feynman diagram of Fig. 1.
We begin by defining the $T$-matrix for the three-nucleon process
depicted in Fig. 1 in terms of the $S$-matrix:
\begin{equation}
  S_{fi} - \delta_{fi}= - (2\pi)^4 i \delta^{(4)}
(\sum p_f - \sum p_i)N_fN_iT_{fi}\, ,
\end{equation}
where $N_f = \prod_{j}(1/2E_j)^{1/2}(2m)^{n_f/2}$ (the index j
labelling all the particles in the final state and $n_f$ being the number
of fermions) and $N_i$ (defined similarly) are the normalization factors
necessary to make $T_{fi}$ the covariant $T$-matrix.  Then the matrix
elements of the momentum
space three-nucleon potential $W$ are given by the non-relativistic
reduction of the three body $S -1$:
\begin{equation}
<\vec{p'}_1\vec{p'}_2\vec{p'}_3|S-1|\vec{p}_1\vec{p}_2\vec{p}_3>_{NR} =
- i (2\pi)\delta(p'_{10} - p'_{20} - p'_{30} -
p_{10} - p_{20} - p_{30})
<\vec{p'}_1\vec{p'}_2\vec{p'}_3|W|\vec{p}_1\vec{p}_2\vec{p}_3> \,\, ,
\end{equation}
so that
\begin{equation}
<\vec{p'_{1}},\vec{p'_{2}},\vec{p'_{3}}| W|\vec{p_{1}},\vec{p_{2}},
\vec{p_{3}}>
 =
(2\pi)^{3} \delta^{3}
   (\vec{p'_{1}}+\vec{p'_{2}}+\vec{p'_{3}}-\vec{p_{1}}-\vec{p_{2}}
   -\vec{p_{3}})
<\vec{p'}_1\vec{p'}_2\vec{p'}_3|T_{NR}|\vec{p}_1\vec{p}_2\vec{p}_3> \,\, .
\label{eq:tnr}
\end{equation}
We note that this choice
for the normalization  corresponds to the use of Dirac spinors
normalized as $\bar{u} u$ = 1 and to momentum eigenstates
such that $<\vec p|\vec{p'}> = {(2 \pi)}^3 \delta (\vec{p}-\vec{p'})$.
Most momentum space three-body codes normalize the momentum space eigenstates
to the Dirac delta function without the factor of $(2\pi)^3$;
this point is discussed rather extensively in Ref. \cite{CG}.
Our choice of convention does not affect the coordinate space formulae
as it is (correctly) absorbed into the Fourier transform, such that
$<\vec r|\vec{r'}> = \delta (\vec{r}-\vec{r'})$.
The full three-nucleon potential W is the sum of three cyclic terms
$W_{1}$, $W_{2}$, and $W_{3}$ where $W_{3}$ has particle 3 in the middle
(active nucleon)
as in Figure 1.  The other two terms are obtained by cyclic permutation
of the indices of the incoming nucleons.

Now we turn our attention to the intermediate state depicted as a blob
in Fig.1.  It is described by the (two-body) $\pi$N scattering amplitude
$\pi^{j} (q) + N (p) \rightarrow \pi^{i} (q') + N (p')$
written in terms of invariant amplitudes $F$ and $B$
which have the the most general isospin decomposition:
\begin{equation}
T^{ij} = -\bar u(p',s')\{ \delta^{ij}(F^{+} - \frac{B^{+}}{4m}
[\rlap/q,\rlap/q'])
     + i \epsilon^{ijk} (F^{-} - \frac{B^{-}}{4m}
[\rlap/q,\rlap/q'])\} u(p,s)
\label{eq:tmat}
\end{equation}
To proceed, one removes the nucleon pole from the four off-shell
crossing symmetric invariants $F^{\pm}(\nu, t, q^2, q'^2)$ and $
B^{\pm}(\nu, t, q^{2}, q'^{2})$. Given the four background $\pi$ N
amplitudes (obtained after the nucleon pole is removed) $\bar
F^{\pm}(\nu, t, q^2, q'^2)$ and $\bar B^{\pm}(\nu, t, q^{2}, q'^{2})$
($\nu= (q+q')\cdot (p+p')/(4m)$, $t=(q-q')^2$), one performs
an expansion in the pion four momenta  $q,q'$.
In this expansion, after terms of the order of $(\mu/m)^2$ ($\mu$
and $m$ stand for the pion and the nucleon mass respectively)
or higher are dropped, only the ($t$-channel) isospin even non-spin flip
$\bar F^{+}$ and the isospin odd spin-flip $\bar B^{-}$ amplitudes
survive \cite{Coon1979,CG}. We will concentrate then on those two. The
non-spin flip even current algebra amplitude is:
\begin{equation}
  \bar F^{+}(\nu ,t,q^{2},q'^{2}) = f(\nu ,t,q^{2},q'^{2}) \frac{\sigma}
    {{f_{\pi}}^2}
      + C^{+}(\nu ,t,q^{2},q'^{2})
 \label{eq:fampli}
\end{equation}
where $\sigma$ is the pion-nucleon $\sigma$ term and
$f_{\pi} \approx 0.93$ MeV.
The double divergence $q'\cdot \bar{M}^+\cdot q/f^2_{\pi}$ of
the background axial vector amplitude denoted by $C^{+}$
contains the higher order $\Delta$ isobar contribution. This amplitude
is given by \cite{ST,Coon1979}
\pagebreak
\begin{eqnarray}
   C^+(\nu,t,q^2,q'^2)
  & = & {2g^{*2}\over 9m^2} {\nu_\Delta\alpha(q'\cdot q)
\over \nu^2_\Delta - \nu^2}
\nonumber \\
   & - &
{g^{*2}(M + m)\over 9 M^2 m} [(M - m)(M+ m)^2
+ (q^2 + q'^2) (2M + m) \nonumber \\
 & - & q'\cdot q(2M - m + (q^2 + q'^2) (M + m)^{-1})] \ ,
 \label{eq:C}
\end{eqnarray}
\noindent where $g^* \approx 1.82 \mu^{-1}$ is the $\pi \Delta$N coupling,
$M=8.825 \mu $ the $\Delta$ mass, $m=6.726 \mu$ the nucleon mass,
$\nu_\Delta = (M - m^2 - q'\cdot q)/2m$ , $\alpha(q'\cdot q)
= (E^2_\pi - q'\cdot q) [(M + m)^2 q'\cdot q] +
(\bar \mu^2 - q'\cdot q) [(M + m)
E_\pi -\frac{1}{2}(\bar \mu^2 + q'\cdot q)]$    with
$E_\pi = (M - m^2 +\bar \mu^2)/2M$
the center of mass energy of the pion at the $\Delta$ resonance.  Here
we have taken ${\bar \mu}^2 = 1/2(q^2 + {q'}^2)$ which reduces to
$\mu^2$ when both pions are on mass shell.

In general, $C^+$ must have the simple form \cite{ST,Coon1979}
\begin{equation}
   C^{+}(\nu ,t,q^{2},q'^{2})=c_{1}{{\nu}^{2}}+c_{2} q \cdot q' +O(q^{4})
\, .
\label{eq:cexpa}
\end{equation}
On the other hand,
the assumed form of the function $f$,
\begin{equation}
     f(\nu ,t,q^{2},q'^{2})=(1-\beta)(\frac{q^{2}+q'^{2}}{{\mu}^{2}}-1)+
\beta (\frac{t}{{\mu}^{2}}-1)
\label{eq:funf}
\end{equation}
(adapted \cite{Coon1979,sigff} for $\pi$N scattering from the $SU(3)$
generalization of the Weinberg low energy expansion for $\pi\pi$
scattering \cite{Wein66})
is such that $\bar F^{+}$ does satisfy the soft pion theorems
\begin{equation}
    \bar F^{+}(0,0,0,0)= - \frac{\sigma}{{f_{\pi}^2}}\,\, ,
\label{eq:softpion}
\end{equation}
\begin{equation}
    \bar F^{+}(0,{\mu}^{2},0,{\mu}^{2})= \bar
F^{+}(0,{\mu}^{2},{\mu}^{2},0) = 0\,\, ,
\label{eq:adler}
\end{equation}
where $C^+$ vanishes identically,
and (with the aid of Eq. (\ref{eq:cexpa})) the constraint
at the (on-shell and measurable)  Cheng-Dashen point:
\begin{equation}
\bar F^{+}(0, 2 {\mu}^{2},{\mu}^{2},{\mu}^{2})= \frac{\sigma}{{f_{\pi}^2}}
  + {\cal O}(q^4)\,.
\label{eq:chengd}
\end{equation}

In contrast to meson-meson scattering, $\beta$ in (\ref{eq:funf}) is not
determined by soft-meson theorems (because the nucleon four momenta
cannot be taken  soft) and  is to be extracted from experiment.
The most recent data analysis from the Karslruhe group \cite{hohlerbook}
makes this extraction
slightly  dependent on t: $\beta$ varies from 0.46
(for $t=0$) to 0.52 (for $t=\mu^2$). This situation is different
from the situation of older data \cite{Coon1979,ST,HJS} which was consistent
with an almost constant $\beta$ value of 0.4 in the same range of $t$.

Neglecting the $\nu^2$ and $q_0$ terms in (\ref{eq:fampli}) because they
are of the order of $(\mu/m)^{2}$ or higher,
 the $\bar F^{+}$ amplitude can be expanded in
the three-vector pion momenta $\vec{q}$ and $\vec{q'}$ as
follows:
\begin{equation}
\bar F^{+}(0,t,q^2,q'^2) = -\frac{\sigma}{{f_{\pi}}^2}
+(\frac{\sigma}{{f_{\pi}}^2} \frac{2 \beta}{\mu^2} - c_2) \vec{q} \cdot
\vec{q'}
-\frac{\sigma}{\mu^2 {f_{\pi}}^2} (\vec{q}\,^2 + \vec{q'}^2)
\label{eq:expan}
\end{equation}

The last equation explicitly exhibits the separation between
the (higher order in $\vec{q}\,^2$) $\Delta$ contribution
--- contained in the $c_2$ term alone ---
and the remaining chiral symmetry breaking terms. In ref.
\cite{Coon1979}
and subsequent discussions of the TM $\pi-\pi$ force,
the $c_2$ and $\beta$ constants in
the coefficient of the $\vec q \cdot \vec {q'}$
term were eliminated in favor
of the on-shell (measurable) quantity $\bar F^{+}(0,\mu^2,\mu^2,
\mu^2)$
\begin{equation}
\bar F^{+}(0,\mu^2,\mu^2,\mu^2)
= (1-\beta)\frac{\sigma}{{f_{\pi}}^2} +\frac{c_2 \mu^2}{2}
\label{eq:fonshell}
\end{equation}
According to Eq. (\ref{eq:expan}), the $\Delta$ term alone does not bring
a structure
in momentum space different from the chiral symmetry breaking term
in $\vec q \cdot \vec {q'}$.
To isolate the $\Delta$ contribution one simply
evaluates the constant $c_2$,
which, from (\ref{eq:cexpa}) (and taking both pions on-mass-shell),
can be done by evaluating
the derivative of (\ref{eq:C})  with respect to $q \cdot q'$, at $t=2
{\mu}^{2}\,\,\,
(q \cdot q' =0)$ and $\nu =0$. The result is
$c_2 = -1.54 {\mu}^{-3}$. This coefficient coincides with twice
$C^{+}$ evaluated at  $t=\mu^2\,\,\, (q \cdot q'=\mu^2/2)$
and $\nu=0$ and compares well with
the value $-1.58 \mu^{-1}$, obtained  for the total contribution from
$C^{+}$ at  $t=0\,\,\, (q \cdot q'= \mu^2)$ and $\nu=0$. The two last
comparisons give an idea of the  negligible importance of the
$O(q^4)$ terms in Eq. (\ref{eq:cexpa}).

{}From the $\pi$N amplitude in conjunction with the $\pi$NN vertices and
pion propagators, one constructs the three body $T_{NR}$,
which according to Eq. (\ref{eq:tnr}) defines the three-body force
represented in Fig. 1 \cite{Coon1979,CG}.
The constant $c_2$ contributes then to the overall
coefficient ``b" that has been used in nuclear calculations
($b= b_{\sigma} + b_{\Delta};\,\, b_{\Delta}=c_2$)
\begin{equation}
b = -\frac{\sigma}{{f_{\pi}}^2} \frac{2 \beta}{\mu^2} + c_2 =
-\frac{2}{\mu^2}(\frac{\sigma}{{f_{\pi}}^2}
- \bar F^{+}(0,\mu^2,\mu^2,\mu^2))
\label{eq:bcoe}
\end{equation}
The value $b=-2.58 \mu^{-3}$, obtained from early experimental results
\cite{HJS}
$\bar F^{+}(0,\mu^2,\mu^2,\mu^2) = -0.16 \mu^{-1}$ and $\sigma/f_{\pi}^2
=1.13 \mu^{-1}$ implies a chiral symmetry breaking
coefficient $b_{\sigma} = b - c_{2} = - 1.04 \mu^{-3}$.
If the updated set of Karlsruhe data is taken instead,
$\bar F^{+}(0,\mu^2,\mu^2,\mu^2)
= - 0.28 \mu^{-1}$ and $\sigma/f_{\pi}^2=1.03
\mu^{-1}$, then $b=-2.62 \mu^{-3}$ and consequently
$b_{\sigma}=-1.08 \mu^{-3}$.
The comparison between the numbers $b_{\Delta}$ and $b_{\sigma}$ exhibits
that the $\Delta$ terms do not dominate the amplitude and any
description which does not obey chiral symmetry breaking \cite{anyofSauer}
is insufficient.

The isolation of the $\Delta$ from the broken chiral symmetry
terms being complete, we have to turn to the backward-propagating
nucleon Born term that is added to the background amplitude, to
generate the three-body force.
That term is calculated by subtracting the forward time-ordered
propagating nucleon term from the covariant nucleon poles.
Such a subtraction was done in refs. \cite{Coon1979,CG}. Dropping
again terms that are of the order of $(\mu/m)^{2}$,
the backward-propagating nucleon term $ F^+_Z$ (``Z-graph") is
\begin{equation}
 F^+_Z = \frac{g^2}{4m^3}(\vec{q}\,^2 + \vec{q'}^2)
\label{eq:bpoles}
\end{equation}
where g is the pseudo-scalar $\pi$NN coupling constant.
Since $ F^+_Z$ shows
the same momentum dependence as the third term in Eq. (\ref{eq:expan}),
the constant $c_z=-g^2/4m^3=-0.15 \mu^{-3}$ contributes,
together with the coefficient of the last term of Eq. (\ref{eq:expan}), to
the overall constant ``c" ($c = c _{\sigma} + c_z$) that multiplies the
$\vec{q}\,^2 + \vec{q'}^2$
term, in the definition of the TM
$\pi-\pi$ three-body force:
\begin{equation}
c = \frac{\sigma}{\mu^2 {f_{\pi}}^2}- \frac{g^2}{4m^3}
 + F'_{\pi NN}(0) \frac{\sigma}{{f_{\pi}}^2}
\label{eq:cterm}
\end{equation}
\noindent We note that the term proportional to $ F'_{\pi NN}(0)$ did not
appear before in Eq. (\ref{eq:expan}).
This term nevertheless is inserted in $c$ because both the
backward propagating part
of the nucleon pole $F^{+}_Z$ and the $\Delta$  couple with the
pion with a (assumed the same) form factor $ F_{\pi NN}(q^2)$ which
is defined as $g(q^2) = g  F_{\pi NN}(q^2)$.
The chiral breaking $\sigma$ term is
a c-number \cite{sigff} and has no intrinsic $q^2$ dependence (although
it is multiplied by $f(\nu, t, q^2, q'^2)$). It is convenient, if not
necessary, however, since part of the amplitude is due to $F^{+}$
and $C^{+}$, to multiply the final amplitude by form factors, dependent
upon $q^2$ and $q'^{2}$. Consequently, the constant term
($\sigma/{f_{\pi}}^2$, labeled ``a" in the literature) attains a
spurious momentum dependence
from the form factors. The term proportional to $ F'_{\pi NN}(0)$ in
Eq. (\ref{eq:cterm}) is inserted to correct for this
spurious momentum dependence to the orders in $q^2$ and $q'^2$
kept in the amplitude.

The value $c = c_{\sigma} + c_z = 1.0 \mu^{-3}$ that has  been used
before changes slightly to $c = 0.91 \mu^{-3}$, with the new
determination of $\sigma/f_{\pi}^2$. From these numbers one concludes
that in the non-spin-flip amplitude amplitude the chiral symmetry
breaking term dominates the relativistic backward-propagating nucleon
term.

The spin-flip current algebra amplitude $\bar B^-$ is simpler to decompose
into the physical contributions. This amplitude is \cite{ST,Coon1979}
\begin{equation}
\bar B^{-} (\nu, t, q^2, q'^2)=
 (F^{V}_1(t) + F^{V}_2(t))\frac{1}{{f_{\pi}}^2} -\frac{g^2}{2 m^2} +
D^{-}(\nu,t,q^2,q'^2)\; ,
\label{eq:bamp}
\end{equation}
where the function $D^{-}$ contains the higher order $\Delta$ contribution
and is given in references \cite{Coon1979,ST,CG}.
After the non-relativistic reduction of the $\pi$N T matrix is done,
in the process of doing the $ \mu /m$ expansion to derive the
three-body-force, we note that $\bar B^-$ is multiplied by
$[\rlap/q, \rlap/q']$ which is already of second order in $\mu/m$,
so we need only keep the first term in the expansion of (\ref{eq:bamp}).
Therefore,
the electromagnetic
isovector form factors are approximated by a constant value
\begin{equation}
F^{V}_1 + F^{V}_2 = 1. + 3.70 + {\cal O}(\frac{\mu^2}{m^2})
\label{eq:formf}
\end{equation}
\noindent and only the constant term ($D_0= 4.87 \mu^{-2}$)
in the expansion of $D^{-}$ is kept.
Taking into account the multiplicative factor $-1/2m$
from Eq. (\ref{eq:tmat}), the $\Delta$ contributes
with a coefficient $d_{\Delta} = D_0/2 m = -0.36 \mu^{-3}$.
The remaining contribution, due to the
two first terms of Eq. (\ref{eq:bamp}), is $d_{ca}= -0.24 \mu^3$.
The sum $d_4 = d_{ca} + d_{\Delta} = 0.60 \mu^{-3}$
defines the  $d_4$ parameter that one encounters in the published equations
for the force.

Finally the $ B^{-}_Z$ contribution from the
backward propagating nucleon pole (``Z-graph") is given to zeroth order
\cite{Coon1979} by
\begin{equation}
 B^-_Z = \frac{g^2}{2m^2} + {\cal O}(\frac{\mu^2}{m^2})
\label{eq:deltab}
\end{equation}
Traditionally then these disparate contributions have been added
into a total ``d" coefficient, $d = d_4 + d_z$, where $d_4$ has already
been defined and $d_z = -B^{-}_Z /2 m$  singles out the pair term
contribution.  The constant $d_z$ has the same value as
the pair term coefficient $d_z = c_z=
-0.15 \mu^{-3}$ of the $F^+$ amplitude.  The total ``d" term coefficient
becomes
$d = d_{ca} + d_{\Delta} + d_z = - 0.75 \mu^{-3}$. Again the numbers
indicate that the pairing term is small relative to the background
amplitude and although the $\Delta$ is important, it is only half of the
``d" coefficient.

Now that we have undone the usual representation of the Tucson-Melbourne
$\pi - \pi$ three-body force, let us close by recalling the form of
the amplitude  $\pi N \rightarrow \pi  N$,
\begin{equation}
      T = T_{Z} + \Delta T  + q'\cdot \bar{M} \cdot q\,\,,
\label{eq:T}
\end{equation}
which best expresses the degree of model dependence of the amplitudes.
Here $T_{Z}$ stands for
the ``Z-graph" contribution, $T_{Z} = T_{B} - T_{FPB}$,
where $T_{B}$ is
the covariant Born term and  $T_{FPB}$ is the forward
propagating (positive energy) nucleon term.  The model independent
 $\Delta T$ is added to the full Born term $T_{B}$
so that to leading order $T_{B} + \Delta T$ satisfies the low
energy theorems (Ward identity constraints) exemplified by
(\ref{eq:softpion})
and (\ref{eq:adler}), and $q'\cdot \bar M \cdot q$ is
a background term for which (isobar) models are necessary. The grouping of
the isobar contributions  into
$\Delta T  + q'\cdot \bar{M} \cdot q $ in Eq.(~\ref{eq:T}) enforces the
largest
degree of model independence of the off-shell amplitude.  This is because
to lowest order (all) isobar contributions are included in $\Delta
T$,  and $q'\cdot \bar{M} \cdot q$ is
constructed to give contributions only in higher order terms.  These
general statements apply equally to the $\rho-\pi$ and the $\rho -
\rho$ amplitudes discussed later on.

What is unique about the $\pi$N $\bar F^+$amplitude is that the
background term $q'\cdot \bar{M} \cdot q$ of Eq. (\ref{eq:cexpa}) has just
the right structure to combine nicely with the momentum dependence
(\ref{eq:funf}) of $\Delta T$ so that the {\em entire} off-shell amplitude
can be obtained directly from data without a specific reference to the
$\Delta$ contributions to either $\Delta T$ or to $q'\cdot \bar{M} \cdot
q$.  This point, while giving us greater confidence in the three-body force
derivation, has lead to considerable frustration among those interested
in three-nucleon forces.  Hence the deconstruction above which is summarized in
the following table.
The table organizes
vertically the coefficients according to the type of momentum and
spin-isospin
structure which ultimately they multiply. Within each row, the
decomposition acording to (\ref{eq:T}), is provided.
The model independent terms labeled $\Delta T$ in the table arise from
current algebra but are of two different types. In the first three rows
(labeled a,b,c) they contain the pion-nucleon sigma term,  an axial
current-axial charge commutator  which is
a measure of chiral symmetry breaking.
The last row (labeled d and originated by $\bar B^{-}$)
contains a $\Delta T$ which is also
a current-current commutator. However  in this case the algebra is closed by
the electromagnetic form factor of the nucleon and has nothing
to do with chiral symmetry breaking.  As we shall see in subsequent
sections, the chiral symmetry breaking contributions to the $\Delta T$
of the $\rho N \rightarrow \pi N$ amplitute are quite small and, of
course, absent altogether in the  $\rho N \rightarrow \rho N$ amplitude
obtained via vector dominance from  $\gamma N \rightarrow \gamma N$.

The specific momentum and
spin-isospin structure of the final three-nucleon force is
provided in momentum space in Ref. \cite{CG}
and coordinate space in Ref. \cite{Joe}, and will not be repeated here.
The
strength constants of the $2\pi$ exchange three-body force are given in units
of the charged pion mass (139.6 MeV).  The potentials, however, use an isospin
formalism instead of charge states so it would seem natural to employ the
SU(2) average pion mass  $(2 m_{\pi^{+}} + m_{\pi^{0}})/3 = 138$ MeV in the
propagators and form factors.

\pagebreak

\baselineskip 0.6cm

\section{The $\rho-\pi$ Potentials}

The working definition of the $\rho-\pi$ three-nucleon potential
is, as in the $\pi-\pi$ case, Eq. (\ref{eq:tnr}) written in section 2.
The three-body transition matrix $T_{NR}$ pictured in figure 2
contains now two different meson propagators together with the $\pi$NN
and the $\rho$NN vertices from the external nucleonic lines of
the diagram, in conjunction with the $\rho N \rightarrow  \pi N$
intermediate amplitude. The amplitude is modeled through vector dominance
from the amplitude $\gamma N \rightarrow \pi N$ which is
obtained by the method of Ward identities \cite{MacMullen}.
Those identities are (single and double) divergence conditions
that determine the terms of the amplitude that are constant
or linear in the exchanged momentum.
Of the several spin-isospin component of the $\rho-\pi$ three nucleon
potential only a few appear to be numerically significant.  They
correspond to the leading terms of the low energy expansion of
$\rho N \rightarrow  \pi N$  truncated to terms of second order, as was
the $\pi - \pi$ force, in powers of the exchanged three-momenta between
the nucleons.  The (t-channel) isotopic spin decomposition of the
amplitude is the same as that of the $\pi$N amplitude of
Eq. (\ref{eq:tmat})
because the $\rho$ is an isotriplet as well as the pion.
Again, the expansion can be rearranged into the form of Eq. (\ref{eq:T})
for an optimal display of its degree of model
independence.
The isospin even  model-independent terms $T^+_Z + \Delta T^+$ nearly
cancel to
leading order because
$\Delta T^+ $ goes to the soft pion
($q \rightarrow 0$) $\rho$ analog Fubini- Furlan-Rosetti limit of pion
photoproduction \cite{MacMullen,BPS}.
The total model independent term in the $\vec{k}\rightarrow 0\, , \; \vec{q}
\rightarrow 0 $ limit takes on the form $1-mg_A(0)/(f_{\pi}g)$ which is
of the order
of few percent (Goldberger-Treiman discrepancy\cite{GT}).  This cancellation
is lessened away from the expansion point
at zero, but then the model-dependent (and $\Delta$ dominated)
background term $q'\cdot \bar{M} \cdot q$ appears to totally dominate
the $\rho N \rightarrow  \pi N$ amplitude.   Therefore, in the isospin
even amplitude, only the $\Delta$-contribution, obtained in a dispersive
sense, is developed into a $\rho-\pi$ potential in this paper.
The situation is just the opposite for the model independent terms
$T^-_Z + \Delta T^-$ of the isospin odd amplitude.  Here the
low energy theorem of pion-photoproduction is due to Kroll and Ruderman
and is simply the ``Z-graph" $T^-_Z$ in pseudoscalar coupling
\cite{MacMullen}.  This leading order (constant) term from the
the $\rho$-analog
Kroll- Ruderman theorem as $ q \rightarrow 0 \, , \, k \rightarrow 0$
should then form the basis of an important part of the $\rho-\pi$
potential.  It must be supplemented, however, by the largest $\rho-\pi$
potential terms arising from the $\Delta$-dominated isospin odd
background.  Potential terms other than the above from $\Delta T^{\pm}$,
including the chiral symmetry breaking terms, appear to be much smaller
than these just mentioned and will be neglected in the following (for
details see Ref. \cite{Ellis}).

We now pick out the terms with one $\rho$ and one $\pi$ from
Figure 2a
which are believed to dominate the expansion and write them in detail.
Those terms correspond to the ``Z-graph" and the seemingly largest terms
from the $\Delta$ isobar.
The contribution from Figure 2b follows from 2a with the substitutions
$1\leftrightarrow 2$ and $\vec{k} \leftrightarrow -\vec{q} $ where
$\vec{k}$ is the three-momentum of the rho and $\vec{q}$ the
three-momentum of the pion.
(In what follows the notation is slightly different from that of
ref. \cite{Ellis} but closer to existing codes and practices).

\subsection{``Kroll-Ruderman" $\rho-\pi$ potential}

The analogue of the Kroll-Ruderman term of pion-photoproduction forms the
basis of a specific local contribution to
the overall $\rho - \pi$ exchange three-body
potential that presents a first order dependence on the exchanged
mesons momentum $\vec k$ and $\vec q$. It is consequently a contribution that
is expected to compete strongly with the delta-isobar terms that
are quadratic in the momentum.

This contribution comes from
the t-channel isospin odd term in the $\rho N \rightarrow \pi N$
amplitude.
  When the pseudoscalar coupling for the $\pi NN$ vertex
is assumed, this term is the
backward-propagating nucleon Born term.
In a form that evidences sequentially from left to right the role of the four
vertices involved in the
diagram, we can write for its momentum space representation
\begin{eqnarray}
      \mompotrp & = &  +(2\pi)^{3} \frac{\delta^{3}
   (\vec{p'_{1}}+\vec{p'_{2}}+\vec{p'_{3}}-\vec{p_{1}}-\vec{p_{2}}
   -\vec{p_{3}})} {(\vec{q}\,^{2} + \mu^{2})(\vec{k}\,^{2}+m^{2}_{\rho}) }
 \nonumber \\
   & &  i \oddiso \: \ksix                                   \nonumber \\
   & & \frac{ g^{2}_{\rho} }{16m^{3}}
   (\fone   + \kapparho \ftwo )\fone    g^{2} F_{\pi NN}^{2}(q^{2})
\label{eq:kr}
\end{eqnarray}
{}From this form it is evident that
the coupling of the $\rho$ meson to the inner ``blob" is
proportional to $\fone$, in accordance with the low
energy theorem
of Kroll and Ruderman. Also, to obtain Eq. (\ref{eq:kr})
we kept the
distinction  between the Dirac and the Pauli couplings of the $\rho$
meson to the outer
nucleon, which appears as the sum $\fone   + \kapparho \ftwo$.
Subsequently, we retain in all equations one form-factor $\fone$ for the Dirac
coupling of the rho to the nucleon and
another form-factor $\ftwo$ for the Pauli coupling.  The latter is multiplied
by
the on-mass-shell $\kappa_{\rho}$  which is taken to have the Hohler and
Peitarinen value 6.6\cite{HP}.  Separate form factors are not usual
in nuclear physics,
where it is common to take the same form for both the charge and
the magnetic momentum form factors. Our choice has the advantage of
being general and flexible to any  type of (independent)
behavior of  the two form factors that experiment may eventually reveal.
The price we have to pay is the introduction of a new parameter
relative to $\ftwo$.

The coupling of
the pion to the nucleon is described by a form-factor $F_{\pi NN}$. Other
form factors are introduced as needed. All form factors are normalized to
unity at the meson on-mass-shell momentum.
As for the
structure, $\vec{\sigma_2} \cdot \vec{q}$ is the non-relativistic coupling
of the $\pi$
to one of the outer nucleon lines, while $\vec{\sigma_1 }\times \vec{k}$
comes from the $\rho$ coupling to the other external nucleon.

In coordinate space the ``Kroll-Ruderman" potential becomes
\begin{eqnarray}
       \corpotrp & = & \delta^{3}(\vec{x'_{1}}- \vec{x_{1}})
                        \delta^{3}(\vec{x'_{2}}- \vec{x_{2}})
                        \delta^{3}(\vec{x'_{3}}- \vec{x_{3}}) \nonumber \\
      & & \frac{-\mu m_{\rho}}{(4\pi)^{2}16m^{3}} g^{2} g^{2}_{\rho}
      \oddiso  (\vec{\sigma_{1}}\times\nabla_{1}\cdot
     \vec{\sigma_{3}})\vec{\sigma_{2}}\cdot\nabla_{2} \nonumber\\
     & &((Z_{DD\rho}(x_{13}) + \kapparho Z_{PD\rho}(x_{13})) Z_{\mu}(x_{23})
\end{eqnarray}
where $\vec{x}_{ij} \equiv \vec{x_{i}} -\vec{x_{j}} $ and
$\nabla_{i}^{j} \equiv \nabla_{i3}^{j} \equiv \partial / \partial x_{i3}
^{j} \equiv
\frac{x_{i3}^{j}}{|x_{i3}|}
d/d x_{i3}$,
($j=1,2,3$ specifies  a given cartesian component and $i=1,2$ a given
nucleon).

The generalized form factor $H(\vec{q}\,^2)$ in each dimensionless
coordinate space
function $Z_{\mu}(x_{ij})$ is typically a product of monopole form
factors: one
for the vertex to the outer nucleon and a second is included in
the coupling of the
meson to the ``blob" of figure 2.
The coordinate space ``propagator" has the generic form

\begin{equation}
Z_{\alpha\beta\mu}(x) = \frac{4\pi}{\mu}
   \int \frac{d^{3}q\,}{(2\pi)^{3}}
\frac{H_{\alpha\beta}(\vec{q}\,^{2})}{\vec{q}\,^{2} + \mu^{2} }
e^{\imath \vec{q}\cdot\vec{x}}     \label{eq:Z}
\end{equation}

where

\begin{equation}
     H_{\alpha\beta}(\vec{q}\,^{2}) = \frac{\Lambda^{2}_{\alpha} - \mu^{2}}
                      { \Lambda^{2}_{\alpha} + \vec{q}\,^{2}}
                      \frac{\Lambda^{2}_{\beta} - \mu^{2}}
                      { \Lambda^{2}_{\beta} + \vec{q}\,^{2}}   .
      \label{eq:H}
\end{equation}
For the exchanged pion  $ H_{\alpha\beta}(\vec{q}\,^{2}) =
F_{\pi NN}^{2}(q^{2})$
so that $\Lambda_{\alpha} = \Lambda_{\beta}$ and the $\alpha,\beta$ indices
are suppressed. The $\rho$ propagator with
subscript ``DD$\rho$" also
indicates $\Lambda_{\alpha} = \Lambda_{\beta}$ with values from the
Dirac coupling $(\fone)^{2}$
and the transformation $\mu \rightarrow m_{\rho}$ but the subscript
``PD$\rho$" indicates
$\Lambda_{\alpha}\neq\Lambda_{\beta}$ with values arising from
the product of
\ftwo and \fone and $\mu \rightarrow m_{\rho}$.

With these form factors and carrying out the indicated derivatives, the
``Kroll-Ruderman" potential becomes
\begin{eqnarray}
       \corpotrp & = &  \frac{-\mu m_{\rho}}{(4\pi)^{2}16m^{3}} g^{2}
                                g^{2}_{\rho}
                       \delta^{3}(\vec{x'_{1}}- \vec{x_{1}})
                        \delta^{3}(\vec{x'_{2}}- \vec{x_{2}})
                        \delta^{3}(\vec{x'_{3}}- \vec{x_{3}}) \nonumber \\
      & & \oddiso (\vec{\sigma_{1}}\times\hat{x}_{13}\cdot
       \vec{\sigma_{3}})\vec{\sigma_{2}}\cdot\hat{x}_{23}       \nonumber\\
      & &  ((X_{1DD}^{\rho}(x_{13}) + \kapparho X_{1PD}^{\rho}(x_{13}))
       X^{\mu}_{1}(x_{23})
\end{eqnarray}
where $X_{1}(x) = Z'(x)$.  Explicitly for
$\Lambda_{\alpha}=\Lambda_{\beta}=\Lambda$
\begin{equation}
      X^{\mu}_{1}(x) = -\mu\left[ G(\mu x) -
      \frac{\Lambda^{2}}{\mu^{2}}G(\Lambda x)  -
      \frac{1}{2} (\frac{\Lambda^{2}}{\mu^{2}} - 1)
       e^{-(\Lambda x)} \right]
\end{equation}
and for $\Lambda_{\alpha}\neq\Lambda_{\beta}$
\begin{equation}
      X^{\mu}_{1\alpha\beta}(x)  =   -\mu \left [ G(\mu x)
      \nonumber \\
         -\left ( \frac{\Lambda_{\beta}}{\mu}\right ) ^{2}
      \left ( \frac{\Lambda^{2}_{\alpha} - \mu^{2}}
       {\Lambda^{2}_{\alpha}-\Lambda^{2}_{\beta}} \right ) G(\Lambda_{\beta}x)
      +\left ( \frac{\Lambda_{\alpha}}{\mu}\right )^{2}
      \left ( \frac{\Lambda^{2}_{\beta} - \mu^{2}}
       {\Lambda^{2}_{\alpha}-\Lambda^{2}_{\beta}} \right ) G(\Lambda_{\alpha}x)
                 \right ]
\end{equation}
where
\begin{equation}
  G(x) = \frac{e^{-x}}{x} \left ( 1 + \frac{1}{x} \right )  .
\end{equation}

Note that the generic $X_{1}(x) = Z'(x)$ has dimensions of mass and the
spin operators
are coupled with unit  radial vectors.  Then the overall constant has
units of MeV and the magnitude is 28.72 MeV.
It should be clear that for rho exchange the generic $X_{1}^{\mu}(x) $
is written as $\mu\rightarrow m_{\rho}$.
This finishes the explicit display of the
``Kroll-Ruderman" term.  It is
expected to be the most important of the model-independent parts of the rho-pi
potential.

\subsection{$\rho\pi$ Potential with $\Delta$ intermediate state }

\subsubsection{ Isospin even amplitude}

Of the eight spin functions the potential decomposes into,
the terms proportional to $1+\kapparho$ are
believed to be the largest because $\kapparho$ has the value 6.6 on the
rho-mass-shell. Those terms which are (t-channel) isospin even are generated
 from
the spin functions  \kfour \  and \kfive. They take the momentum space form
(see Eq. 2.19a of ref. \cite{Ellis}.):
\begin{eqnarray}
      \mompotrpdelplus & = &  -(2\pi)^{3} \frac{\delta^{3}
   (\vec{p'_{1}}+\vec{p'_{2}}+\vec{p'_{3}}-\vec{p_{1}}-\vec{p_{2}}
   -\vec{p_{3}})} {(\vec{q}\,^{2} + \mu^{2})(\vec{k}\,^{2}+m^{2}_{\rho}) }
 \nonumber \\
   & &   \eveniso \: (\vec{k}\cdot\vec{k}\kfour - \vec{k}\cdot\vec{q}\kfive)
                                   \nonumber \\
   & & \frac{1}{48m^{5}}
    g_{\rho}(\fone+\kapparho \ftwo )   \bigGstark \nonumber \\
   & &\frac{m}{M} \frac{5M-m}{M-m}g F_{\pi NN}(q^{2})(m \gstar)
   \label{eq:deltaplusrp}
\end{eqnarray}
where $M$ is the mass of the Delta and $m$ is the nucleon mass.
The two terms can be combined as
$(\rhoNN )\cdot(\vec{q}\times\vec{k})\piNN $ which manifests the
$\rho$NN coupling to ``outer" nucleon 1 and the $\pi$NN coupling to
``outer" nucleon 2.  On the other hand, we note that the second term of
Eq.~\ref{eq:deltaplusrp} has the same structure
($\kfive \vec{k}\cdot\vec{q}$ )
as the ``b" term of the $2\pi$ exchange potential.

In coordinate space the potential becomes

\begin{eqnarray}
       \corpotrpdelplus & = & \delta^{3}(\vec{x'_{1}}- \vec{x_{1}})
                        \delta^{3}(\vec{x'_{2}}- \vec{x_{2}})
                        \delta^{3}(\vec{x'_{3}}- \vec{x_{3}}) \nonumber \\
      & & \frac{-\mu m_{\rho}}{(4\pi)^{2}
       48Mm^{4}}   g g_{\rho} \frac{5M-m}{M-m}G^{*}_{M\rho} (mg^{*})
      \nonumber\\ & & \eveniso
   [  \vec{\nabla_{1}}\cdot\vec{\nabla_{1}}
\vec{\sigma_{1}}\cdot\vec{\nabla_{2}}\vec{\sigma_{2}}\cdot\vec{\nabla_{2}}
 - \vec{\nabla_{1}}\cdot\vec{\nabla_{2}}
\vec{\sigma_{1}}\cdot\vec{\nabla_{1}}\vec{\sigma_{2}}\cdot\vec{\nabla_{2}} ]
   \nonumber\\ & &
       ((Z_{DG\rho}(x_{13}) + \kapparho
       Z_{PG\rho}(x_{13})) Z_{N\Delta\mu}(x_{23})  \label{eq:twoderiv}
\end{eqnarray}
where subscripts $``DG\rho"$ refer to the generalized form factor arising
from the product of \fone and the $\rho\Delta N$ form factor \Frhondel,
$``PG\rho"$ to the analogous product $\ftwo \Frhondel$, and
$``N\Delta\mu"$ to $\Fpinn\Fpindel$.

The second derivatives of Eq. (\ref{eq:twoderiv}) are most
conveniently carried out with the aid of the following basic identity
\begin{equation}
      \partial_i\partial_j Z(x) = \frac{1}{3}[\delta_{ij}Y_{2}(x) +
       (3\hat{x}_{i}\hat{x}_{j} -\delta_{ij})X_{2}(x)]
\label{eq:partial}
\end{equation}
($\hat{x_i} =\vec{x_i} \ |x_i|$)
where
$X_{2}(x) = Z''(x) - 1/x Z'(x)$, and $Y_{2}(x) = Z''(x)+ 2/x Z'(x)$.
Explicitly they are
\begin{eqnarray}
      X^{\mu}_{2}(x)& =& \mu^{2}\left[ F(\mu x) -
      \frac{\Lambda^{3}}{\mu^{3}}F(\Lambda x)
      -  \frac{1}{2}\frac{\Lambda}{\mu}\Lambda x
      (\frac{\Lambda^{2}}{\mu^{2}} - 1) G(\Lambda x) \right] ,\nonumber \\
      Y^{\mu}_{2}(x) & = &\mu^{2}\left[ \frac{e^{-\mu x}}{\mu x} -
      \frac{\Lambda^{3}}{\mu^{3}} \frac{e^{-\Lambda x}}{\Lambda x}
      -  \frac{1}{2}\frac{\Lambda}{\mu}\Lambda x
      (\frac{\Lambda^{2}}{\mu^{2}} - 1)(1-\frac{2}{\Lambda x})
      \frac{e^{-\Lambda x}}{\Lambda x}
       \right] .
\end{eqnarray}
  Because
\begin{equation}
   F(x) = \frac{e^{-x}}{x}(1 + \frac{3}{x} + \frac{3}{x^{2}} )
\end{equation}
for the seldom needed $\Lambda_{\alpha}=\Lambda_{\beta}=\Lambda$ we can
make the identifications
$X^{\mu}_{2}(x) = \mu^{2}T(x)$ and
 $Y^{\mu}_{2}(x) = \mu^{2}Y(x)$ familiar from the Argonne-Urbana two nucleon
force expressions \cite{Joe}.  In the present case, however,
 $\Lambda_{\alpha}\neq
\Lambda_{\beta}$ for each propagator and we must employ the longer forms

\begin{eqnarray}
      X^{\mu}_{2\alpha\beta}(x)& =& \mu^{2}\left[ F(\mu x)
      - \frac{\Lambda^{3}_{\beta}}{\mu^{3}}
      (\frac{\Lambda^{2}_{\alpha}-\mu^{2}}
      {\Lambda^{2}_{\alpha}-\Lambda^{2}_{\beta}}) F(\Lambda_{\beta} x)
      + \frac{\Lambda^{3}_{\alpha}}{\mu^{3}}
      (\frac{\Lambda^{2}_{\beta}-\mu^{2}}
      {\Lambda^{2}_{\alpha}-\Lambda^{2}_{\beta}})
           F(\Lambda_{\alpha} x)  \right] ,\nonumber \\
      Y^{\mu}_{2\alpha\beta}(x)& = & \mu^{2}\left[ \frac{e^{-\mu x}}{\mu x}
      - \frac{\Lambda^{3}_{\beta}}{\mu^{3}}
      (\frac{\Lambda^{2}_{\alpha}-\mu^{2}}
      {\Lambda^{2}_{\alpha}-\Lambda^{2}_{\beta}})
        \frac{e^{-\Lambda_{\beta} x}} {\Lambda_{\beta} x}
      + \frac{\Lambda^{3}_{\alpha}}{\mu^{3}}
      (\frac{\Lambda^{2}_{\beta}-\mu^{2}}
      {\Lambda^{2}_{\alpha}-\Lambda^{2}_{\beta}})
        \frac{e^{-\Lambda_{\alpha} x}} {\Lambda_{\alpha} x}
            \right] .
\end{eqnarray}

Before displaying the final potential in r-space
let us introduce the ``tilde" notation
for the combination $1+\kapparho$ which appears so often. For example,
the $\rho$ propagator in (\ref{eq:twoderiv})
which has two form factors
in the $\rho NN$ coupling and only one in the $\rho\Delta N$,
takes on the compact form
\begin{equation}
    \tilde{Z}_{NG\rho}(x) = Z_{DG\rho}(x) + \kapparho Z_{PG\rho}(x).
\end{equation}
Since differentiation is a linear operator we can write, for example,
\begin{equation}
    \tilde{X}^{\rho}_{2}(x)\equiv \tilde{X}^{\rho}_{2NG}(x)
             = X_{2DG}^{\rho}(x) + \kapparho X_{2PG}^{\rho}(x).
\end{equation}
to shorten the following equations.


\newcommand{\ytwo}[2] {\mbox {$Y_{2}^{#1}(x_{#2})   $}}
\newcommand{\xtwo}[2] {\mbox {$X_{2}^{#1}(x_{#2})   $}}

\newcommand{\yone}[2] {\mbox {$Y_{1}^{#1}(x_{#2})   $}}
\newcommand{\xone}[2] {\mbox {$X_{1}^{#1}(x_{#2})   $}}
\newcommand{\spinspin} {\mbox { $\vec{\sigma_{1}}\cdot\vec{\sigma_{2}}   $}}
\newcommand{\tensor}[1] {\mbox {$3\vec{\sigma_{1}}\cdot\hat{x}_{#1}
                                 \vec{\sigma_{2}}\cdot\hat{x}_{#1}
              -  \vec{\sigma_{1}}\cdot\vec{\sigma_{2}}   $}}
\newcommand{\sonetwo}[1] {\mbox { $S_{12}(\hat{x}_{#1}) $}}

Using Eq. (\ref{eq:partial}) the first term of Eq. (\ref{eq:twoderiv})
 (from $ \kfour\vec{k}\cdot\vec{k}$ ) becomes
\begin{eqnarray}
       {\corpotrpdelplus}_1 & = & \delta^{3}(\vec{x'_{1}}- \vec{x_{1}})
                        \delta^{3}(\vec{x'_{2}}- \vec{x_{2}})
                        \delta^{3}(\vec{x'_{3}}- \vec{x_{3}}) \nonumber \\
      & & \frac{-\mu m_{\rho}}{(4\pi)^{2}
       48Mm^{4}}   g g_{\rho} \frac{5M-m}{M-m}G^{*}_{M\rho} (mg^{*})
      \nonumber\\ & &
      \eveniso [\frac{\spinspin}{3}\tilde{Y}_{2}^{\rho}(x_{13})\ytwo{\mu}{23}
      \nonumber \\
     & & +  \frac{\sonetwo{23}}{3}\tilde{Y}_{2}^{\rho}(x_{13})\xtwo{\mu}{23}]
\label{eq:first}
\end{eqnarray}
 The second term of Eq. (\ref{eq:twoderiv}) has the same structure
($\kfive \vec{k}\cdot\vec{q}$ )
as the ``b" term of the $2\pi$ exchange potential and is

\begin{eqnarray}
       {\corpotrpdelplus}_2 & = & \delta^{3}(\vec{x'_{1}}- \vec{x_{1}})
                        \delta^{3}(\vec{x'_{2}}- \vec{x_{2}})
                        \delta^{3}(\vec{x'_{3}}- \vec{x_{3}}) \nonumber \\
      & & \frac{+\mu m_{\rho}}{(4\pi)^{2}
       48Mm^{4}}   g g_{\rho} \frac{5M-m}{M-m}G^{*}_{M\rho} (mg^{*})
      \nonumber\\ & &
      \eveniso \frac{1}{9}[\spinspin \tilde{Y}_{2}^{\rho}(x_{13})
                                      \ytwo{\mu}{23} \nonumber \\
     & & +  \sonetwo{13}\tilde{X}_{2}^{\rho}(x_{13})\ytwo{\mu}{23}
         +  \sonetwo{23}\tilde{Y}_{2}^{\rho}(x_{13})\xtwo{\mu}{23}  \nonumber
\\
     & & + (9 \sigma_{1}\cdot\hat{x}_{13} \sigma_{2}\cdot\hat{x}_{23}
             \hat{x}_{13}\cdot\hat{x}_{23} - \sonetwo{13} -\sonetwo{23}
 \nonumber \\
     & &   -\spinspin )\tilde{X}_{2}^{\rho}(x_{13})
            \xtwo{\mu}{23}  ] \label{eq:bterm}
\end{eqnarray}

The sum of these two terms agrees with the corresponding equation (2.23) of
\cite{Ellis}.  Identification of Eq. (\ref{eq:bterm}) with the coordinate
space ``b" term as given in ref. \cite{Joe}
can be made by
remembering that $\sonetwo{13} = \tensor{13} $ and noting that
$ [Y_{2}(x) - X_{2}(x)]/3 = Z'(x)/x  $.   The overall coordinate space constant
(including the powers of meson masses
in the definitions of $X^{\mu}_{2}(x)$ and  $Y^{\mu}_{2}(x)$) for both terms
(\ref{eq:first}) and (\ref{eq:bterm}) is 524.28 MeV.
 This number is huge because the $\rho N\Delta$ coupling constant
\bigGstar is (rather artifically) defined on the $\rho$ mass shell but is
in reality
obtained from the experimental
 $G^{*}_{M\gamma}(k^{2}=0) = 14.7$.  That is
$G^{*}_{M\gamma}(k^{2}=0) = G^*_{M\rho} F_{\rho N \Delta}(k^2=0)$.
Solving for $G^*_{M\rho}$, one finds $G^*_{M\rho} = G^{*}_{M\gamma}(k^{2}=0)
\frac{\Lambda^2}{\Lambda^2 - m^2_{\rho}}$ where $\Lambda = 5.8 \mu$ and
$m_{\rho} \approx 5.6 \mu$.   The product $\bigGstark$ is employed in the
subthreshold region for space-like momentum transfer, where it is much smaller.

\subsubsection{Isospin odd amplitude}

Those largest terms which are (t-channel) isospin odd multiply
the spin functions  of Eqs. (2.13f) and (2.23g) in Ref. \cite{Ellis}:
$m^2 {\cal K}_6 = \ksix$ and $m^4{\cal K}_7 = \kseven$.  The
particular combination containing the delta-nucleon mass difference takes
the form  $m^4{\cal K}_7 - \vec{q}\cdot\vec{k}m^2{\cal K}_6 =
\kseven - \vec{q}\cdot\vec{k} \ksix$. This can be rewritten as
$i \rhoNN\cdot[(\vec{\sigma_{3}}\times\vec{q})\times\vec{k}]\piNN$ to show
the spin structure connected with the several vertices most clearly,
or finally  as
$-i(\vec{\sigma_{3}}\cdot\vec{q}\times\vec{k}\vec{\sigma_{1}}\cdot\vec{k}
\piNN  - \vec{\sigma_{1}}\times\vec{\sigma_{3}}\cdot\vec{q} (\vec{k})^2
\piNN)$.
The
third form is convenient for the coordinate space manipulations and has
appeared before from non-relativistic transition potential derivations
\cite{Mart,Gari}.  We will quote the momentum space potential with the
last form of the spin functions

\begin{eqnarray}
      \mompotrpdelminus & = &  -(2\pi)^{3} \frac{\delta^{3}
   (\vec{p'_{1}}+\vec{p'_{2}}+\vec{p'_{3}}-\vec{p_{1}}-\vec{p_{2}}
   -\vec{p_{3}})} {(\vec{q}\,^{2} + \mu^{2})(\vec{k}\,^{2}+m^{2}_{\rho})}
\:  i \oddiso  \nonumber \\    & &
i[\vec{\sigma_{3}}\cdot\vec{q}\times\vec{k}\vec{\sigma_{1}}\cdot\vec{k}
\piNN  - \vec{\sigma_{1}}\times\vec{\sigma_{3}}\cdot\vec{q} (\vec{k})^2
\piNN]   \nonumber \\
& & \frac{1}{96m^{5}}
g_{\rho}(\fone+\kapparho \ftwo )   \bigGstark \nonumber \\    & &
\frac{m}{M}\frac{M+m}{M-m}g F_{\pi NN}(q^{2})(m \gstar)
\end{eqnarray}

One can readily see that the first term has the same spin structure
$\vec{\sigma_{3}}\cdot\vec{q}\times\vec{k}\vec{\sigma_{1}}\cdot\vec{k}
\piNN$ as the  ``d" term of the 2$\pi$ exchange potential.
In coordinate space the isospin odd potential becomes

\begin{eqnarray}
       \corpotrpdelminus & = & \delta^{3}(\vec{x'_{1}}- \vec{x_{1}})
                        \delta^{3}(\vec{x'_{2}}- \vec{x_{2}})
                        \delta^{3}(\vec{x'_{3}}- \vec{x_{3}}) \nonumber \\
      & & \frac{\mu m_{\rho}}{(4\pi)^{2}
       96Mm^{4}}   g g_{\rho} \frac{M+m}{M-m}G^{*}_{M\rho} (mg^{*})\oddiso
      \nonumber\\ & &
   [ \vec{\sigma_{3}}\cdot \vec{\nabla_{2}}\times\vec{\nabla_{1}}
\vec{\sigma_{1}}\cdot\vec{\nabla_{1}}\vec{\sigma_{2}}\cdot\vec{\nabla_{2}}
 - \vec{\sigma_{1}}\times\vec{\sigma_{3}}\cdot\vec{\nabla_{2}}
\vec{\nabla_{1}}\cdot\vec{\nabla_{1}}\vec{\sigma_{2}}\cdot\vec{\nabla_{2}} ]
   \nonumber\\ & &
       ((Z_{DG\rho}(x_{13}) + \kapparho
       Z_{PG\rho}(x_{13})) Z_{N\Delta\mu}(x_{23})  \label{eq:x}
\end{eqnarray}

After the derivatives are carried out, the first term of Eq. (\ref{eq:x}),
analogous to the ``d" term of the 2$\pi$ exchange potential, takes the
form:
\begin{eqnarray}
       {\corpotrpdelminus}_1 & = & \delta^{3}(\vec{x'_{1}}- \vec{x_{1}})
                        \delta^{3}(\vec{x'_{2}}- \vec{x_{2}})
                        \delta^{3}(\vec{x'_{3}}- \vec{x_{3}}) \nonumber \\
      & & \frac{\mu m_{\rho}}{(4\pi)^{2}
       96Mm^{4}}   g g_{\rho} \frac{M+m}{M-m}G^{*}_{M\rho} (mg^{*})
      \nonumber\\ & & \oddiso
[(\vec{\sigma_{3}}\cdot\hat{x}_{23}\times\hat{x}_{13})
 (\vec{\sigma_{1}}\cdot\hat{x}_{13})(\vec{\sigma_{2}}\cdot\hat{x}_{23})
  \nonumber \\ & &
        \tilde{X}_{2}^{\rho}(x_{13})\xtwo{\mu}{23}  \nonumber \\
 & &
 +(\vec{\sigma_{1}}\cdot\hat{x}_{13})
(\vec{\sigma_{2}}\cdot\hat{x}_{13}\times\vec{\sigma_{3}})\frac{1}{x_{23}}
       \tilde{X}_{2}^{\rho}(x_{13})\xone{\mu}{23}   \nonumber \\
 & &
+(\vec{\sigma_{2}}\cdot\hat{x}_{23})
(\vec{\sigma_{3}}\cdot\hat{x}_{23}\times\vec{\sigma_{1}})\frac{1}{x_{13}}
       \tilde{X}_{1}^{\rho}(x_{13})\xtwo{\mu}{23}   \nonumber \\
& &
 + (\vec{\sigma_{2}}\cdot\vec{\sigma_{1}}\times\vec{\sigma_{3}})
\frac{1}{x_{13}}\tilde{X}_{1}^{\rho}(x_{13})\frac{1}{x_{23}}\xone{\mu}{23}]
	\label{eq:dterm}
\end{eqnarray}

The second term of Eq. (\ref{eq:x}) does not have a counterpart in
the 2$\pi$ exchange potential but instead takes the simple form:

\begin{eqnarray}
       {\corpotrpdelminus}_2 & = & \delta^{3}(\vec{x'_{1}}- \vec{x_{1}})
                        \delta^{3}(\vec{x'_{2}}- \vec{x_{2}})
                        \delta^{3}(\vec{x'_{3}}- \vec{x_{3}}) \nonumber \\
      & & \frac{\mu m_{\rho}}{(4\pi)^{2}
       96Mm^{4}}   g g_{\rho} \frac{M+m}{M-m}G^{*}_{M\rho} (mg^{*})
      \nonumber\\ & & \oddiso
[+(\vec{\sigma_{2}}\cdot\hat{x}_{23})
(\vec{\sigma_{3}}\cdot\vec{\sigma_{1}}\times\hat{x}_{23})
       \tilde{Y}_{2}^{\rho}(x_{13})\xtwo{\mu}{23}   \nonumber \\
& &
 + (\vec{\sigma_{2}}\cdot\vec{\sigma_{3}}\times\vec{\sigma_{1}})
\frac{1}{x_{23}}\tilde{Y}_{2}^{\rho}(x_{13})\xone{\mu}{23}]
	\label{eq:extra}
\end{eqnarray}
The overall numerical constant in both (\ref{eq:dterm}) and
(\ref{eq:extra}) is 109 MeV, about a factor of five smaller than that of
the isospin even term ( for this reason this term was not included in the
nuclear matter calculation of ref. \cite{Ellis}; in a trinucleon calculation
however it may turn out to be meaningful).
A model three-nucleon potential derived from the
transition potential approach would have an exact ratio of 4:1 for the
isospin even versus isospin odd contributions.

One local term proportional to
$i[\vec{\sigma_{1}}\times\vec{k}\cdot\vec{\sigma_{3}}\piNN]  \vec k
\cdot\vec {k}$ remains from the expansion of the covariant expression
for the amplitude with an intermediate $\Delta$.  The overall numerical
constant is   about a factor of ten smaller than that of
(\ref{eq:dterm}) and (\ref{eq:extra}) so we will not consider it
further.  It requires a third derivative of $Z$ or first derivative of
$Y_2$  in coordinate space anyway, which eventually suppresses it
further.



The total rho-pi contribution is obtained from
${\cal W}^{\rho\pi}(3A)$ by first adding the contribution from
${\cal W}^{\rho\pi}(3B) = ({\cal W}^{\rho\pi}(3A):1\leftrightarrow 2)$
and then taking cyclic permutations of three ``active" nucleons .

\pagebreak

\section{The $\rho-\rho$ Potentials}

The definition of Eq. (\ref{eq:tnr}) is once more the starting point.
The three-body $T_{NR}$ for the Feynman diagram of figure 3 involves
the $\rho N \rightarrow \rho N$ amplitude, together with two $\rho$
meson propagators and $\rho$NN vertices. The calculation of
the latter amplitude is obtained by generalizing the Thirring theorem for
Compton scattering to the case of $\rho$ mesons, through the
current-field identity, under the assumption of vector dominance.
The longitudinal character of the $\rho$ has however to be
considered as well as the fact that it carries isospin. To do
that one proceeds as did Beg \cite{Beg} in his analysis of Compton
scattering of isovector photons. The dominant term in the background
(free of the nucleon pole term) Beg amplitude arises from
the t channel  $\rho$ pole ($ 3 \rho$ coupling) and from a
direct $\rho\rho$NN contact (seagull) term.
Those contributions constitute the $\Delta T$ part of
Eq. (\ref{eq:T}). The $\Delta$ term, contributing to the
quadratic terms in the momenta, cannot be fixed by any
divergence condition and is introduced, explicitely separated
from the remaining amplitude, in a model dependent way.
The following subsections will turn to the detailed form
of all these components of the $\rho$N amplitude of different (physical)
origin.

\subsection{Model-independent potentials from low-energy theorems}

The apparent dominant terms in the model-independent parts of this potential
are
odd  in the t-channel isospin.  The term displayed below results from
the coupling of transverse (T) $\rho$-mesons to the low-energy amplitude
constructed by Beg  and has been labeled the ``Beg" potential in the
previous nuclear matter calculations. Similarly to the Kroll-Ruderman term,
the ``Beg" term is linear in each of the meson momentum, and it turns
out to be
important when comparing to the $\Delta$ isobar contributions.
In momentum space it takes the
form

\begin{eqnarray}
      \mompotrr & = &  +(2\pi)^{3} \frac{\delta^{3}
   (\vec{p'_{1}}+\vec{p'_{2}}+\vec{p'_{3}}-\vec{p_{1}}-\vec{p_{2}}
   -\vec{p_{3}})} {(\vec{k}\,^{2}+m^{2}_{\rho})(\vec{k'}\,^{2}+m^{2}_{\rho}) }
\oddiso  \nonumber \\
   & &  [(\vec{\sigma}_{1}\times\vec{k})\cdot(\vec{\sigma}_{2}\times
\vec{k'})\times\vec{\sigma}_{3}]\, F_{1 \rho}^{2}(0)
                    (1 + \frac{F_{2V}(0)}{F_{1 \rho}^{2}(0)}) \nonumber    \\
   & & \frac{g^{4}_{\rho}}{64m^{3}}
   (\fone + \kapparho \ftwo )(\fonep + \kapparho \ftwop ) \nonumber \\
\end{eqnarray}
where  $F_{2V}(0)=\kappa_V=3.7$  is the isovector anomalous moment
of the nucleon.  While the term independent of $F_{2V}$ in
$F_{1 \rho}^{2}(0)(1 + \frac{F_{2V}(0)}{F_{1 \rho}^{2}(0)})$
arises from the backward-propagating
nucleon pole term (or pair term), the term in $F_{2V}(0)$ represents
a direct $\rho\rho$NN contact (seagull) term in the underlying
$\rho\rm N \rightarrow\rho\rm N$ amplitude.  In principal the second term
should be distinguished by a form factor  of a different character
($F_{1\rho}((k-k')^2)$ than the previous terms of the
$\rho\pi$ potential. In practice,  this component of the $\rho\rho$
potential has so far been calculated in the approximation of no
form factors on the ``active" nucleon and therefore includes only
the $\rho$NN form factor at the external nucleons.  This approximation
is similar to the approximation made in the ``d" term of the $2\pi$
force. In that case, only the constant term of the current commutator
expansion $F_{1V}(t) + F_{2V}(t) \approx F_{1V}(0) + F_{2V}(0) + {\cal
O} (q^2) \approx 1 + 3.70 $ was retained (see Eq. (18)).

In coordinate space the ``Beg" potential has a slightly different form
from the previous ones displayed, because the propagator Z of Eq.
has only a single monopole in the H function of Eq. (\ref{eq:H})
and (as in the ``Kroll-Ruderman" term) only the
first derivative is needed.  With the definitions
\begin{equation}
      Z'_{1\mu} \equiv      X^{\mu}_{11}(x) = -\mu\left[ G(\mu x) -
      \frac{\Lambda^{2}}{\mu^{2}}G(\Lambda x)   \right],
\end{equation}
where the added subscript 1 is meant to remind about the single formfactor,
the coordinate space potential becomes

\begin{eqnarray}
       \corpotrr & = & \delta^{3}(\vec{x'_{1}}- \vec{x_{1}})
                        \delta^{3}(\vec{x'_{2}}- \vec{x_{2}})
                        \delta^{3}(\vec{x'_{3}}- \vec{x_{3}}) \nonumber \\
      & & \frac{ m^{2}_{\rho}}{(4\pi)^{2}64m^{3}}  g^{4}_{\rho}
           F_{1 \rho}^{2}(0) (1 + \frac{F_{2V}(0)}{F_{1 \rho}^{2}(0)})
   \nonumber\\
      & &  \oddiso
       (\vec{\sigma_{1}}\times\nabla_{1})\cdot(\vec{\sigma_{2}}\times
        \nabla_{2})\times\vec{\sigma_{3}} \nonumber\\
     & & \tilde{Z}_{1m_{\rho}}(x_{13})\tilde{Z}_{1m_{\rho}}(x_{23})
\end{eqnarray}
or

\begin{eqnarray}
       \corpotrr & = &  \frac{ m^{2}_{\rho}}{(4\pi)^{2}64m^{3}}  g^{4}_{\rho}
           F_{1 \rho}^{2}(0) (1 + \frac{F_{2V}(0)}{F_{1 \rho}^{2}(0)})
   \nonumber\\
     & &                  \delta^{3}(\vec{x'_{1}}- \vec{x_{1}})
                        \delta^{3}(\vec{x'_{2}}- \vec{x_{2}})
                        \delta^{3}(\vec{x'_{3}}- \vec{x_{3}}) \nonumber \\
     & &   \oddiso  (\vec{\sigma_{1}}\times\hat{x}_{13})\cdot
    (\vec{\sigma_{2}}\times\hat{x}_{23})\times\vec{\sigma_{3}} \nonumber\\
      & &  \tilde{X}_{11}^{\rho}(x_{13})\tilde{X}^{\rho}_{11}(x_{23})
\end{eqnarray}
\subsection{$\rho\rho$ Potential with $\Delta$ intermediate state}

There are two local potentials which correspond to the coupling of
transverse $\rho$-mesons to the $\Delta$ pole.  We introduce the notation

\begin{equation}
\lambda_{M} \approx -\frac{3}{2m(M+m)}\bigGstar
\end{equation}
for the low-momentum transfer $\rho\Delta$N vertex.  The two local terms
correspond to t-channel isospin even and odd amplitudes.  We take them
up in turn.

\subsubsection{Isospin even amplitude}

The form of the term generated by this amplitude agrees
with the form arising from a transition potential construction
and reads in momentum space

\begin{eqnarray}
      \mompotrrplusdel & = &  -(2\pi)^{3} \frac{\delta^{3}
   (\vec{p'_{1}}+\vec{p'_{2}}+\vec{p'_{3}}-\vec{p_{1}}-\vec{p_{2}}
   -\vec{p_{3}})} {(\vec{k}\,^{2}+m^{2}_{\rho})(\vec{k'}\,^{2}+m^{2}_{\rho}) }
 \nonumber \\
   & &   \eveniso \: \rhocrosscross    \nonumber    \\
   & & \frac{g^{2}_{\rho} \lambda_{M}^{2}} {18(M - m)}
      (\fone + \kapparho \ftwo )F_{\rho N\Delta}(k^{2})\,  \nonumber  \\
   & &   (\fonep + \kapparho \ftwop ) F_{\rho N\Delta}(k'^{2})\,
\end{eqnarray}

which, after Fourier transform becomes

\newcommand{\corrhocrosscross} {\mbox { $
((\vec{\sigma_{1}}\times\vec{\nabla_1})\times\vec{\nabla_1})\cdot
((\vec{\sigma_{2}}\times\vec{\nabla_2})\times\vec{\nabla_2})  $}}

\begin{eqnarray}
 \corpotrrplusdel & = & \delta^{3}(\vec{x'_{1}}- \vec{x_{1}})
                        \delta^{3}(\vec{x'_{2}}- \vec{x_{2}})
                        \delta^{3}(\vec{x'_{3}}- \vec{x_{3}}) \nonumber \\
        & & - \frac{m^2_{\rho}}{(4\pi)^2}
      \frac{g^{2}_{\rho} \lambda_{M}^{2}} {18(M - m)}
        \eveniso \:\:  \nonumber\\
   & &    [\corrhocrosscross]   \nonumber \\
   & &   (Z_{DG\rho}(x_{13}) + \kapparho
           Z_{PG\rho}(x_{13}))   (Z_{DG\rho}(x_{23}) +
          \kapparho Z_{PG\rho}(x_{23})).
\end{eqnarray}

\newcommand{\tytwo}[1] {\mbox {$\tilde{Y}_{2}^{\rho}(x_{#1})   $}}
\newcommand{\txtwo}[1] {\mbox {$\tilde{X}_{2}^{\rho}(x_{#1})   $}}

\newcommand{\tyone}[1] {\mbox {$\tilde{Y}_{1}^{\rho}(x_{#1})   $}}
\newcommand{\txone}[1] {\mbox {$\tilde{X}_{1}^{\rho}(x_{#1})   $}}

Carrying out the derivatives one obtains the final form of the matrix
element in r-space

\begin{eqnarray}
 \corpotrrplusdel & = & \delta^{3}(\vec{x'_{1}}- \vec{x_{1}})
                        \delta^{3}(\vec{x'_{2}}- \vec{x_{2}})
                        \delta^{3}(\vec{x'_{3}}- \vec{x_{3}}) \nonumber \\
      & & - \frac{m^2_{\rho}}{(4\pi)^2}
      \frac{g^{2}_{\rho} \lambda_{M}^{2}} {18(M - m)}
        \eveniso \frac{1}{9}\:\:  \nonumber  \\
 & & ([4\spinspin\tytwo{13}\tytwo{23} - 2\sonetwo{13}\txtwo{13}\tytwo{23}
    \nonumber \\  & &
-2\sonetwo{23}\tytwo{13}\txtwo{23}]  +  [9\hat{x}_{13}\cdot\hat{x}_{23}
\vec{\sigma_1}\cdot\hat{x}_{13}\vec{\sigma_2}\cdot\hat{x}_{23}
    \nonumber \\  & &
 -\sonetwo{13} - \sonetwo{23} - \spinspin ] \txtwo{13} \txtwo{23} ] ).
\end{eqnarray}

\subsubsection{Isospin odd amplitude}

There is a relative factor of $-1/4$ between isospin odd and isospin
even in this case. This is precisely the relative factor obtained from a
transition potential approach. The reason why the TM force ends up with
the same factor lies in the fact that the  $k\cdot C \cdot k'$
contributions to the $\rho\rm N \rightarrow\rho\rm N$ amplitude were
modeled by $\Delta$-poles in the s- and u-channels.

\newcommand{\corrhotriplecross} {\mbox { $
((\vec{\sigma_{1}}\times\vec{\nabla_1})\times\vec{\nabla_1})\times
((\vec{\sigma_{2}}\times\vec{\nabla_2})\times\vec{\nabla_2})  $}}

\begin{eqnarray}
      \mompotrrminusdel & = &  +(2\pi)^{3} \frac{\delta^{3}
   (\vec{p'_{1}}+\vec{p'_{2}}+\vec{p'_{3}}-\vec{p_{1}}-\vec{p_{2}}
   -\vec{p_{3}})} {(\vec{k}\,^{2}+m^{2}_{\rho})(\vec{k'}\,^{2}+m^{2}_{\rho}) }
 \nonumber \\
   & &    \frac{1}{4} \oddiso \:\: \vec{\sigma}_{3}\cdot [\rhotriplecross]
  \nonumber \\
   & &  \frac{g^{2}_{\rho} \lambda_{M}^{2}} {18(M - m)}
         (\fone + \kapparho \ftwo )F_{\rho N\Delta}(k^{2})\, \nonumber \\
   & &   (\fonep + \kapparho \ftwop ) F_{\rho N\Delta}(k'^{2})\,
\end{eqnarray}

In coordinate space the equation becomes:

\begin{eqnarray}
 \corpotrrminusdel & = & \delta^{3}(\vec{x'_{1}}- \vec{x_{1}})
                        \delta^{3}(\vec{x'_{2}}- \vec{x_{2}})
                        \delta^{3}(\vec{x'_{3}}- \vec{x_{3}}) \nonumber \\
      & & \frac{+m^2_{\rho}}{(4\pi)^2}
      \frac{g^{2}_{\rho} \lambda_{M}^{2}} {18(M - m)}  \frac{1}{4}
        \oddiso \:\:  \nonumber\\
   & &   \vec{\sigma}_{3}\cdot [\corrhotriplecross]   \nonumber \\
   & &   (Z_{DG\rho}(x_{13}) + \kapparho
           Z_{PG\rho}(x_{13}))   (Z_{DG\rho}(x_{23}) +
          \kapparho Z_{PG\rho}(x_{23}))
\end{eqnarray}

We note that the expansion of the vector products in the previous equation
leads to four terms, one of which is identical to the ``d" term of the
$\pi-\pi$ force.

After carrying out the derivatives we find finally:

\begin{eqnarray}
 \corpotrrminusdel & = & \delta^{3}(\vec{x'_{1}}- \vec{x_{1}})
                        \delta^{3}(\vec{x'_{2}}- \vec{x_{2}})
                        \delta^{3}(\vec{x'_{3}}- \vec{x_{3}}) \nonumber \\
      & & \frac{+m^2_{\rho}}{(4\pi)^2}
      \frac{g^{2}_{\rho} \lambda_{M}^{2}} {18(M - m)}  \frac{1}{4} \oddiso
      \nonumber\\
   & &   [ (\vec{\sigma_2} \cdot\vec{\sigma_3}\times\vec{\sigma_1})\tytwo{13}
	\tytwo{23}  \nonumber \\ & &
      -(\vec{\sigma_3}\cdot\vec{\sigma_1}\times\hat{x}_{23})
   (\vec{\sigma_2}\cdot\hat{x}_{23}\tytwo{13}\txtwo{23}
    \nonumber\\  & &
   -(\vec{\sigma_1}\cdot\vec{\sigma_2}\times\vec{\sigma_3})\tytwo{13}
\frac{\txone{23}}{x_{23}}
 \nonumber \\  & &
 - (\vec{\sigma_3}\cdot\hat{x}_{13}\times\vec{\sigma_2})
(\vec{\sigma_1}\cdot\hat{x}_{13})\txtwo{13}\tytwo{23}  \nonumber \\  & &
   -  (\vec{\sigma_3} \cdot\vec{\sigma_1}\times\vec{\sigma_2})
    \frac{\txone{13}}{x_{13}}\tytwo{23}]  \nonumber \\  & &
- [(\vec{\sigma_3}\cdot\hat{x}_{23}\times\hat{x}_{13})(\vec{\sigma_1}\cdot
\hat{x}_{13})(\vec{\sigma_2}\cdot\hat{x}_{23})\txtwo{13}\txtwo{23}
\nonumber \\  & &
 -
(\vec{\sigma_1}\cdot\hat{x}_{13})(\vec{\sigma_2}\cdot\hat{x}_{13}\times
\vec{\sigma_3})\frac{1}{x_{23}}\txtwo{13}\txone{23} \nonumber \\  & &
- (\vec{\sigma_2}\cdot\hat{x}_{23})(\vec{\sigma_1}\cdot\vec{\sigma_3}
\times\hat{x}_{23})\frac{\txone{13}}{x_{13}}\txtwo{23}
\nonumber \\  & &
-(\vec{\sigma_2}\cdot\vec{\sigma_1}\times\sigma_3)
  \frac{\txone{13}}{x_{13}}\frac{\txone{23}}{x_{23}}]
\end{eqnarray}

where the four last terms correspond to the ``d" term analog piece of
the matrix element.

\section{Recommended choice of parameters}

The set of numerical values for the coupling constants and  the monopole
form-factor regulator masses are compiled in Table II.  The monopole
formfactors are defined in Eqs. (23) and (24).

This set differs from table A2 of ref. \cite{Ellis} only in the value for
the $\pi$N$\Delta$ coupling constant, which should be $g^{*}=1.82
\mu^{-1}$,
instead of the larger value of that table. Only this value is consistent with
the $\pi$N$ \Delta$ coupling constant implicit in the 2$\pi$ exchange
three-body
force (see Appendix A of Ref. \cite{Coon1979}) and with the value
obtained by the Karlsruhe group \cite{hohlerbook} from $\pi$N scattering
analysis.  They used a fixed-t dispersion relation for the invariant
amplitudes in order to determine the parameters of an ``effective pole"
which simulates $\Delta$-exchange at energies near and below threshold.
The coupling constant so determined corresponds to the dimensionless
coupling $f^{*}$ (${f^{*}}^2/ 4\pi = 0.26$) used by Martzolff et al
\cite{Mart}, and differs from the Hannover group choice ${f^{*}}^2 /
4\pi = 0.35$ for their transition potential \cite{HSS}.  The latter value
is derived from the $\Delta$-width, using the interaction term of the
Lagrangian and ``should not be used in application of pole term formulas
near or below threshold" \cite{hohlerquote}. This latter value is 40\%
higher and therefore emphasizes the effects of the forward propagating
isobar in the transition potential picture compared to the covariant isobar
contribution to  $\pi$N scattering.

The decision for the prescribed values of the remaining parameters
was justified in ref. \cite{Ellis}. Briefly, we recall here
that,
\begin{itemize}
\item{ the coupling constants concerning the $\pi$
come from low energy $\pi-N$ scattering data.  The value of $g=13.4$ has
been challenged in recent years by partial wave analyses of
nucleon-nucleon scattering \cite{stoks}.  Much discussion has ensued on
this still unsettled point.}

\item{ the cut-off value for the $\pi N\Delta$ form factor is taken to
be the same as $\pi$NN, which is suggested by data from neutrino-nucleon
scattering ($p + \nu \rightarrow \Delta ^{++} + \mu^{-}$).}

\item{ for the $\rho$ meson, the choices made on the parameters were
determined through vector meson dominance models (which ties the needed
hadronic $\rho\Delta$N coupling to the $\gamma\Delta$N vertex) that are
believed to be valid for processes involving low four-momentum transfer.
 We follow the experimental analyses of the  $\gamma\Delta$N vertex and
use a monopole form factor with cutoff mass  $\Lambda_{\rho N \Delta} =
5.8 \mu$ to approximate this vertex. As for the $\rho$NN vertex, it has
two couplings: a direct Dirac coupling and the Pauli coupling to the
anomalous magnetic moment of the nucleon. In the Tucson-Melbourne
prescription, the ratio of the Pauli to Dirac coupling is allowed to
vary with the four-momentum of the $\rho$. This choice  appears to
accommodate reasonably both the vector dominance value of
$\kappa_{V}(k^2=0)=3.7$ and the on-mass-shell value of
$\kappa_{\rho}(k^2={m_{\rho}}^2) = 6.6$ \cite{HP}. The {\em two} cutoff
masses $\Lambda_{Dirac} = 12 \mu$ and $\Lambda_{Pauli} = 7.4 \mu$ are
then determined from nucleon electromagnetic form factor data.  Somewhat
later, a group led by Gari \cite{Gariff} made a detailed fit to the
electromagnetic form factors of the nucleon with the ansatz suggested
already in \cite{Ellis} and came to a similar conclusion as
\cite{Ellis}. What was not noticed by the authors of \cite{Ellis} and
presumably of \cite{Gariff} was that the $k^2$ dependence of the Dirac and
Pauli rho form factors away from the rho mass shell were also shown and
even parameterized for spacelike momentum transfer by H\"{o}hler and
Pietarinen in \cite{HP}.  Indeed, the ratio of Pauli to Dirac coupling
from the latter  analysis does not vary much with $k^2$ and is 6.0 at
$k^2=0$ \cite{hohlerpc}.  It turns out that this form factor result for
the $\rho$NN vertex is quite compatible with the data for the
electromagnetic isovector nucleon form factors \cite{hohlerff}.}
\end{itemize}

We next discuss the most varied aspect of the TM force: the fact that
each user feels free to substitute his or her own choice for the
recommended cut-off value of the monopole $\pi NN$ form factor. Its
recommended value ($\Lambda=5.8 \mu \approx 810$ MeV) is inspired by the
$5 \%$ Goldberger-Treiman discrepancy (which, measuring chiral symmetry
breaking, is consistent with a $3 \%$ decrease in the $\pi NN$ form
factor from $q^2={m_{\pi}}^2$ to $q^2=0$ \cite{GT}. The present
understanding of the variation of the other quantities in the
Goldberger-Treiman relation from the chiral limit to the chirally broken
``real world", imply an ``error bar" of $\pm 200$ MeV for the $\Lambda$
determination. This makes our choice in good agreement with the OPEP
tail in many contemporary NN potentials: $\Lambda \approx 980$ MeV for
Argonne V14 \cite{pionmom}, $\Lambda = 800$ MeV in the present day
evolution of the Bonn OBEP potentials \cite{HHT}, and $\Lambda = 800$
MeV for both $\pi$ and $\rho$ in the Bochum potential \cite{Gari}. The
$\pi$NN form factor $\Lambda \approx 1300$ MeV of the full Bonn
potential \cite{fullBonn} has always been inconsistent with evidence other
than partial wave analysis of the NN system \cite{GT};
this difficulty is expected to be
overcome if additional diagrams with correlated $\rho-\pi$ exchange is
included into the full model \cite{Holinde}.  On the other hand, the
cutoff mass $\Lambda_0 = 965$ MeV in the  exponential form factor of
Nijmegen potentials \cite{Nijmegenpot} corresponds to   $\Lambda
=\sqrt{2} \Lambda_0 \approx 1365$ MeV for a monopole form factor
\cite{stoks}, rather like that of the full Bonn potential.

We should note that the current controversy over the value of the
on-shell $\pi$NN coupling $g$ affects the cutoff mass determination from
the Goldberger-Treiman discrepancy $\Delta_{GT} = 1-mg_A(0)/(f_{\pi}g) \approx
0.05 $;
a smaller coupling constant implies a smaller discrepancy and a larger
cutoff mass.  But other analyses which isolate one-pion exchange via a
Regge analysis of the charge exchange data $d \sigma / dt |_{np}$ minus
$d \sigma / dt |_{\bar{p}p}$ also find a $3 \%$ decrease in the $\pi NN$
form factor from $q^2={m_{\pi}}^2$ to $q^2=0$ \cite{GT}, indicating a
certain robustness in the recommended value of the cutoff mass.

The recent consensus on a low mass cutoff for pion exchange highlights
the point already emphasized by the Hokkaido group \cite{Sapporo} and,
in the modern context, by the S\~{a}o Paulo group \cite{contact}.  The
contact terms (those proportional to a coordinate-space
$\delta$-function and its derivatives) are spread out with increasing
importance as $\Lambda$ becomes smaller and the size of the nucleon
grows.  The dominant attitude so far was to consider that these contact
terms, bringing the nucleon structure signature, should not be included
in potential models. In contrast, the low cutoff mass of the $\pi$NN and
$\rho$NN form factors of the Bochum NN potential, means that {\em all}
mesons more massive than the $\rho$ are subsumed into contact terms.
This reformulation of the traditional OBEP  can still yield a
satisfactory description of the NN data, although a $\chi^{2}$ analysis
is needed for a definite judgment.  The nice feature of the Bochum NN
potential is that the small two-pion exchange term is highly suppressed,
cut down by four powers of a rapidly varying cutoff function and
partially replaced by the contact terms. Weinberg obtains similar
contact terms in his NN potential from a presumed four-fermion
interaction \cite{Weinberg}.

We finish this section by returning to the subject of the $\rho$NN
formfactor.  We have developed the formulas in momentum and coordinate
space for two separate Dirac and Pauli formfactors of the monopole type.
Clearly it is easy to specialize these general formulas to a single
formfactor and then the choice of parameters is between
the vector dominance value of $\kappa_{V}(k^2=0)=3.7$ expected for a
narrow resonance and the
on-mass-shell value of $\kappa_{\rho}(k^2={m_{\rho}}^2) =6.6$ obtained from
a dispersive analysis of $\pi$N scattering.
{}From our current understanding of the the $\rho$NN vertex,
a better but more difficult to implement form factor might be those
(Dirac and Pauli) of Eq. 4.3 in \cite{HP} which goes into a monopole for
high $k^2$ but has a more complicated structure at small $k^2$ near 0.
We suggest that future users of three-body forces might concentrate
on the $\rho$NN vertex rather than on the $\pi$NN vertex, where the form
factor of the Tucson-Melbourne three-nucleon potentials is not only
determined by particle and nucleon-nucleon data but, in addition,  is rather
consistent with most of the realistic two-nucleon potentials.

\section {Numerical Results}

It is well-known that calculations with many  recent ``realistic" two-nucleon
potentials (Argonne V14, Nijmegen potential or the folded
diagram version of the full Bonn potential) produce underbinding
of the the trinucleon bound state \cite{Elba}. However, the same interactions,
when taken in conjunction with the TM two-pion exchange three-body force,
provide too much binding. There are  overlapping and
retarded pion exchange graphs not discussed in this paper which yield
non-local terms \cite{CF}. These have
not been numerically evaluated yet, but are expected to be less important
than the (already small) nucleon-antinucleon pair terms of the 2$\pi$-TM
potential and henceforth expected not to alter the binding defect problem.

To test the idea that the three-nucleon potentials of the two-meson
exchange structure extended to include $\rho$-exchange act against the
overbinding effect of the 2$\pi$-TM potential, we estimated perturbative
contributions to the binding energy of the triton from the extended TM
force described in detail in the previous sections.

The calculation was done using a wavefunction in the coordinate-space
obtained by solving the Faddeev equations with the
Malfliet-Tjon I-III potential \cite{MTpot}.  We present the results in
Table III,
where the individual contributions from the different physical processes
considered in the force are singled out.
The contributions from the $\pi-\pi$ force are slightly different from
preliminary results presented before \cite{Elba} because in this paper
we use the expansion coefficients of Table I. \cite{corr}  We stress
that the wavefunction
used is  ``semi-realistic" at most (no NN p or d waves are included in the
Malfliet-Tjon I-III potential).
The past history of results obtained with the $\pi-\pi$ force showed
us already that any calculation
with less than 18 three-body channels has to be considered with
caution.
The numbers given here were produced merely in the spirit of
having a quick and hopefully qualitative information on the importance
of the $\rho$ exchange in the three-nucleon force. A much more
sophisticated calculation, which is out of the scope of this paper
whose aim is restricted to
provide the force to be used,
will be reported \cite{Stadler2,Stadler3} soon.

{}From table III we conclude first that the $\rho$ exchange does indeed modify
the effect of the $\pi-\pi$ force and that the effect goes in the (expected)
direction of less binding.
Since it has been fashionable to
take the derived strengths of a nuclear force seriously, but to consider the
meson-nucleon form factor cutoffs as adjustable parameters, we show
in the columns labeled ``Martzolff" a set of calculations which
keep the strength constants of the
Tucson-Melbourne three-body force but use the much heavier
cutoff masses from Ref. \cite{Mart}
(we had to convert however their monopole parameterization of the product
of two vertices to our dipole convention to find $\Lambda_{\pi} = 10.6 \mu$
and  $\Lambda_{\rho} = 13.4 \mu$ for {\em all} couplings to N's or
$\Delta$'s). We note then that although the contributions
of the $\pi-\pi$ are, as is well known,
strongly cutoff dependent,
the inclusion of the $\pi$-$\rho$  force supresses, at least in
this triton model, much of that cutoff dependence.
Finally, the results obtained for the TM force show a satisfying pattern
of decreasing effect with increased mass of the exchanged meson. The same
pattern does not exist in the pointlike
nucleons (large cutoff masses) calculation of the rightmost columns.
In any event, the addition
of three-body forces due to $\rho$ exchange, especially the $\rho-\pi$ exchange
force, does indeed counter
the too strong
attraction of the $\pi$-$\pi$ and seems to stabilize the total effect under
variations in form factors.  We await future tests of these tentative
conclusions with some eagerness.

\begin{center}
Acknowledgements
\end{center}
M. T. P. wants to thank the Theory Group at CEBAF for the hospitality granted
during
her stay.
This work was supported in part by the National Science Foundation and
by Junta Nacional de Investiga\c c\~ao Cient\'\i fica (JNICT).  We are
especially
grateful for a NATO Collaborative Research Grant which enabled us to
undertake and finish this project.

\pagebreak

\newpage

\begin{table}[hbt]
\begin{center}
\begin{tabular}{ccrcr}
        & $T_{Z} \,$  &\multicolumn{1}{c}{$\Delta T$}   &
$ q'\cdot \bar{M}_{\Delta} \cdot q \,$  &\multicolumn{1}{c}{Total} \\ \hline
$\mu$a   &     0  &    +1.03     &     0   & +1.03    \\

$\mu^{3}$b &     0 &    -1.08    &    -1.54  & -2.62    \\
$\mu^{3}$c &    -0.15 &   +1.06    &      0   & +0.91   \\
$\mu^{3}$d &    -0.15 &  -0.24     &   -0.36      & -0.75 \\
\label{tab1}
\end{tabular}
\caption{Expansion coefficients of the $\pi$N amplitude used in the
$\pi-\pi$ force. Units of charged pion masses.
$T_{Z}$ is the "Z-graph" contribution,
$\Delta T$ is the model independent part needed to satisfy the low
energy theorems,
$q'\cdot \bar M \cdot q$ is
the background term for which (isobar) models are necessary.}
\end{center}
\end{table}

\newpage
\begin{table}[hbt]
\begin{center}
\begin{tabular}{llr}
for the $\pi$& & \\
\hline
       & $g$                       &13.4\\
       & $g^{*}$                 & 1.8 $\mu ^{-1}$\\
       & $\Lambda_{\pi NN}$      &  5.8 $\mu$\\
       & $\Lambda_{\pi N\Delta}$ & 5.8 $\mu$\\
for the $\rho$& & \\
\hline
       &$ g_{\rho}$                      & 5.3\\
       &${G_{M \rho}}^{*}$               &191.1\\
       & $\Lambda_{\rho NN}$ (Dirac)     & 12 $\mu$ \\
       & $\Lambda_{\rho NN}$ (Pauli)     & 7.4 $\mu$  \\
       & $\Lambda_{\rho N\Delta}$        & 5.8 $\mu$ \\
\label{tab2}
\end{tabular}
\end{center}
\caption{ Couplings labeled by a generic $g$ and cut-off masses
labeled by $ \Lambda $.  See Ref. 3 for complete definitions.}
\end{table}

\newpage

\begin{table}[hbt]
\begin{center}
\begin{tabular}{llrrrrrrrr}
\multicolumn{2}{c}{Exchanged mesons}&
\multicolumn{4}{c}{Tucson-Melbourne form factors} &
\multicolumn{4}{c}{``Martzolff" form factors} \\
\hline
  &  &$T_z$ &$\Delta T$&$q'\cdot \bar{M}_{\Delta}\cdot q$&Total
     &$T_z$ &$\Delta T$&$q'\cdot \bar{M}_{\Delta}\cdot q$& Total \\
$\pi-\pi$ & & & & & & & & & \\
         &$T^+$             & 0.055&-0.409    &  0.090 &-0.264
                            & 0.011&-0.363    & -0.257 &-0.609  \\
         &$T^-$             &-0.032&-0.052    & -0.077 &-0.161
                            &-0.043&-0.070    & -0.105 &-0.218\\
         &Total             & 0.023&-0.461    &  0.013 &-0.425
                            &-0.032&-0.433    & -0.362 &-0.827\\
\hline
$\rho-\pi$&  &  &  &  &  &   &  &  \\
  &$T^{+}$& & &-0.022 & -0.022
          & & &0.261 &0.261\\
  &$T^{-}$&0.215 & &0.090  &0.090
          &0.342 & &0.340 &0.340\\
  & Total  &0.215& &0.068  &0.283
           &0.342& &0.601&0.943\\
\hline
$\rho-\rho$&    &       &        &  &  &   &  &  &  \\
  &$T^+$&       &       &-0.005  &-0.005
        &       &       &-0.101  &-0.101\\
  &$T^-$&0.001  &0.002  &-0.007  &-0.004
        &0.005  &0.019  &-0.247  &-0.223\\
 &Total &0.001  &0.002  & -0.012 &-0.009
        &0.005  &0.019  &-0.348  &-0.324\\
\hline
TOTAL &  &  &  &  & -0.151  &  &  &  &-0.208  \\
\label{tab3}
\end{tabular}
\caption{ The three-body force contributions (in MeV) to the triton
binding energy. The contributions are arranged according to Eq.
(20). Distinction between the isospin even and odd contributions is also shown.
The calculation was made perturbatively with a (three-channel)
Malfliet-Tjon I-III model wave function.}
\end{center}
\end{table}

\newpage

\begin{center}
FIGURE CAPTIONS\\
\end{center}

Figure 1 --- Diagram for the $\pi-\pi$ force\\
\vspace{0.5cm}

Figure 2 --- Diagrams for the $\pi-\rho$ force\\
\vspace{0.5cm}

Figure 3 --- Diagram for the $\rho-\rho$ force\\
\vspace{0.5cm}


\begin{thebibliography}{999}
\bibitem{corr} In addition, a mistake in the sign of the
(miniscule) isospin odd $\rho - \rho$ amplitude contribution from $ T_Z
+ \Delta T$ (`` Beg") has been corrected since Ref. \cite{Elba}.
\end{thebibliography}
\end{document}